\pdfoutput=1
\documentclass[prd,nofootinbib,twocolumn,superscriptaddress,longbibliography]{revtex4-1}
\usepackage{enumerate}
\usepackage{amsfonts,amsmath,amssymb}
\usepackage[backref=none]{hyperref}
\usepackage{graphicx,subfigure,bm}
\usepackage{color}
\usepackage{footnote}
\usepackage{tikz}
\usepackage[utf8]{inputenc}
\usepackage{textcomp}
\usepackage{array}
\usepackage{fixltx2e}
\usepackage{diagbox}
\usepackage{caption}
\newcommand*{\rom}[1]{\expandafter\@slowromancap\romannumeral #1@}

\definecolor{MyDarkBlue}{rgb}{0,0.1,0.7}

\hypersetup{pdfborder={0 0 0},colorlinks,breaklinks=true,
  urlcolor={MyDarkBlue},citecolor={MyDarkBlue},linkcolor={MyDarkBlue}}

\def\equationautorefname~#1\null{Eq.(#1)\null}

\newcommand{\dif}{{\rm d}}
\def\apj{Astrophysical Journal}

\def\apss{Astrophysics and Space Science}
\def\jcap{Journal of Cosmology and Astroparticle Physics}
\def\prd{Physical Review D}
\def\prc{Physical Review C}
\def\nat{Nature}

\def\aap{Astronomy and Astrophysics}
\def\araa{Annual Reviews of Astronomy and Astrophysics}

\def\physrep{Physics Reports}
\begin{document}

\title{Constraint on energy-momentum squared gravity from neutron stars\\ and its cosmological implications}

\author{\"{O}zg\"{u}r Akarsu}
\email{akarsuo@itu.edu.tr}
\affiliation{Department of Physics, {\.I}stanbul Technical University, Maslak 34469 {\.I}stanbul, Turkey}

\author{John D. Barrow}
\email{J.D.Barrow@damtp.cam.ac.uk}
\affiliation{DAMTP, Centre for Mathematical Sciences, University of Cambridge, Wilberforce Road, Cambridge CB3 0WA, U.K.}

\author{Sercan {\c C}{\i}k{\i}nto{\u g}lu}
\email{cikintoglus@itu.edu.tr}
\affiliation{Department of Physics, {\.I}stanbul Technical University, Maslak 34469 {\.I}stanbul, Turkey}

\author{K. Yavuz Ek{\c s}i}
\email{eksi@itu.edu.tr}
\affiliation{Department of Physics, {\.I}stanbul Technical University, Maslak 34469 {\.I}stanbul, Turkey}

\author{Nihan Kat{\i}rc{\i}}
\email{nihan.katirci@itu.edu.tr}
\affiliation{Department of Physics, {\.I}stanbul Technical University, Maslak 34469 {\.I}stanbul, Turkey}

\begin{abstract}
Deviations from the predictions of general relativity due to energy-momentum squared gravity (EMSG) are expected to become pronounced in the high density cores of neutron stars. We derive the hydrostatic equilibrium equations in EMSG and solve them numerically to obtain the neutron star mass-radius relations for four different realistic equations of state. We use the existing observational measurements of the masses and radii of neutron stars to constrain the free parameter, $\alpha ,$ that characterizes the coupling between matter and spacetime in EMSG. We show that $-10^{-38}\,\mathrm{cm^{3}/erg}<\alpha <+10^{-37}\,\mathrm{cm^{3}/erg}$. Under this constraint, we discuss what contributions EMSG can provide to the physics of neutron stars, in particular, their relevance to the so called  \textit{hyperon puzzle} in neutron stars. We also discuss how EMSG alters the dynamics of the early universe from the predictions of the standard cosmological model. We show that EMSG leaves the standard cosmology safely unaltered back to $t\sim 10^{-4}$ seconds at which the energy density of the universe is $\sim 10^{34}\,\mathrm{erg\,cm^{-3}}$.

\end{abstract}
\date{\today}

%\keywords{}
\maketitle
\section{Introduction}
\label{section:intro}

Einstein's general theory of relativity (GR) agrees with all tests in the solar system to a precision of $10^{-5}$ \cite{wil14}. The discovery of the late-time acceleration of the universe (see \cite{cal09} for a review) led
to a reintroduction of the cosmological constant ($\Lambda$), a possible energy density of the quantum vacuum energy density in the universe. The huge discrepancy between its value calculated from quantum field theory and that required to explain the accelerating cosmic expansion, however, led to an extensive search for alternative explanations for the accelerating expansion. A broad avenue followed by many cosmological studies is to introduce modifications to GR (see \cite{sot10,def10,noj11,cap11} for a review) which can lead to accelerating cosmological solutions \cite{car+04,cog+08}. These should reduce to GR in the weak gravity field limit in order to be consistent with the classical solar system tests. Yet, there are models of gravity which make similar predictions in the weak field limit, but deviate from GR in the strong field regime \cite{psa08,psa09,ber+15}.

Almost all of these modifications to GR focus on generalizing the gravitational Lagrangian away from the linear function of scalar curvature, $\mathcal{R}$, responsible for the Einstein tensor in Einstein's equations. On the other hand, it is possible to consider generalizing the form of the matter Lagrangian in a nonlinear way, for instance, to some analytic function of the scalar $T^{2}=T_{\mu \nu }T^{\mu \nu }$ formed from the energy-momentum tensor (EMT), $T_{\mu \nu }$, of the matter stresses, as first discussed in \cite{kat14}. Such a generalization of GR includes new type of contributions of the material stress to the right-hand side of the Einstein equations without invoking some new forms of fluid stress, such as bulk viscosity or scalar fields. A particular example of this type of generalizations in the form $F(\mathcal{R},T^{2})=\mathcal{R}+\alpha T^{2}$, dubbed as energy-momentum squared gravity (EMSG), was proposed in \cite{ros16}, and very recently a more general one, in the form $F(\mathcal{R},T^{2})=\mathcal{R}+\alpha(T^{2})^{n}$, dubbed as energy-momentum powered gravity (EMPG), was proposed in \cite{aka17,barrow17}. In EMPG model, the case $n>1/2$ (EMSG corresponds to $n=1$) may be effective at high energy densities, e.g., relevant to early universe and dense compact astrophysical objects, while the case $n<1/2$ may be effective at low energy densities, e.g., relevant to dynamics of the late universe. Namely, for $n>1/2$, it can replace the initial singularity with an initial bounce and avoid spatial anisotropy from dominating the universe about the initial singularity \cite{barrow17}, and, for $n<1/2$, it can lead to $\Lambda $CDM type cosmology without invoking $\Lambda $ when $n=0$, and $w$CDM like cosmologies without invoking a dark energy source when $n\simeq 0$ \cite{aka17}. The reader is referred to Ref. \cite{barrow17} (and the references therein) for a detailed discussion motivating the type of generalization of GR including higher-order contributions to the right-hand side of the Einstein equations, where the material stresses appear.

In this paper, we focus on EMSG, which leads to quadratic contributions to gravity from matter terms, which then can be effective at high energy densities and pressures. The loop quantum gravity and braneworld scenarios contribute such new quadratic terms by replacing $\rho$ by $\rho (1\pm O(\rho ^{2}))$ in Einstein's equations, where the negative contribution is from loop quantum gravity \cite{LQC11} and the positive is from braneworld scenarios \cite{BraneW1}. EMSG can affect the cosmological dynamics significantly at energy densities higher than a certain energy density depending on the value of its free parameter $\alpha $. Hence, to probe the energy densities where EMSG would lead to significant deviations from GR and have consequences for the initial singularity, inflation, big bang nucleosynthesis, or detailed structure of the microwave background power spectrum, it is necessary to constrain $\alpha $ observationally. EMSG is equivalent to GR in vacuum and hence its effects could be apparent only inside matter sources. Indeed, as we will discuss below, the strongest constraints on $\alpha$ can be obtained from neutron stars (NSs) and then the cosmological implications of EMSG under these constraints can also be investigated.

Investigation of EMSG could also be interesting for the possibility of addressing some problems in the physics of NSs such as the so called \textit{hyperon puzzle}\footnote{Hyperons are baryons containing at least one strange quark. These particles are not stable on Earth, decaying to nucleons through weak interactions, but are stabilized at the degenerate cores of NS \cite{amb60}.}. The hyperonization of matter leads to the softening of the equation of state (EoS) which then reduces the maximum mass of the NS to $\sim 1.4M_{\odot }$. Although the appearance of hyperons seems unavoidable \cite{cha16}, the predicted maximum mass is at odds with the measured masses of $\simeq 2M_{\odot }$ \cite{dem+10,ant+13}. A recently proposed way to alleviate the \textit{hyperon puzzle} is to modify gravity at strong gravity field \cite{ast+14}. Consideration of compact stars in braneworld scenarios, leading to Einstein's equations reminiscent of EMSG for $\alpha >0$, is not new \cite{Maartens01}, and very recently, it was claimed that \textit{hyperon puzzle} can be addressed, on account of non-linear contributions of matter stress to the right-hand side of the Einstein's equations, in the Randall-Sundrum type-II braneworld in \cite{Lugones17} and in the braneworld within the Eddington-inspired Born-Infeld gravity in \cite{Prasetyo18}. In EMSG, the effective stiffening in the right-hand side of the Einstein equations, due to the new contributions of matter stresses under some conditions, may compensate the softening of the matter EoS due to the hyperonization and so enhance the maximum masses of NSs with hyperons. The investigation of this possibility, however, is not straightforward and it is necessary to employ the numerical solutions of structure of NSs and their mass-radius relations for various realistic EoS parametrizations.

It is noteworthy that, because the EoS of NSs has not been constrained by terrestrial experiments, there are several EoS parameterizations and hence one may expect degeneracy between a modified gravity and different EoS parameterizations, yet as it was shown in \cite{ara+11,eks+14}, it is possible to use NSs to constrain the order of magnitude of the free parameter/s of a modified theory, yet the constraint would still be much tighter than what could be obtained from solar system tests.

There are two measures of the strength of gravity: compactness ($\eta \equiv2GM/Rc^{2}$) and spacetime curvature ($\xi \equiv 4\sqrt{3}GM/R^{3}c^{2}$) where $M$ is the mass scale and $R$ the length scale of a system with energy density $\rho $ \cite{psa08}. Cosmological studies probe gravity at large compactness $\eta =8\pi G\rho R^{2}/3c^{2}\sim 1$ while the curvature is very weak ($\xi <<1$) because of the large length scales involved.

The black holes are the most compact objects, but the vacuum solutions around black holes in most modified theories of gravity are similar to GR \citep{Haw72,bek72, bek+78,sot+12,psa+08} except in Chern-Simons gravity, making any discrimination between these models hard to observe by probing black holes despite $\eta=1$. This leaves NSs as the best sources to constrain modified theories of gravity. Indeed, the compactness and curvature of a typical NS of mass $M=1.4\,M_{\odot }$ and radius $R=10\,\mathrm{km}$, respectively, are $10^{5}$ and $10^{14}$ times larger than the values probed in solar system tests \citep{ded03} but they still are in an unexplored regime in the bulk of the NS \citep{eks+14}. There is considerable effort \citep{coo+10,ara+11,cap+11,pan+11,del+12,ast+13,yaz+14,gan+14,ast+14,ast+15a,ast+15b,cap+16,ara+17,ast+17,don17} to study the mass-radius relation of NSs in modified theories of gravity. In this paper we seek to determine the form of the mass-radius relations for NSs in the EMSG theories in order to determine whether this theory can survive confrontation with observations of NS environments for a selection of four realistic equations of state. The EMSG theory we investigate is characterized by a single coupling constant, whose numerical value turns out to be severely constrained by the structure of NSs. In the next section we introduce the structure of the EMSG theory we are investigating before deriving the equations of hydrostatic equilibrium for NSs in Sec. \ref{sec:hydrostatics}. In Sec. \ref{sec:method}, we briefly describe the numerical method employed to determine the mass-radius relations for the NSs in Sec. \ref{sec:results}. These lead to a tight constraint upon the defining coupling constant in the EMSG theory considered here and in Sec. \ref{cosmo_analysis} we use that constraint to discuss the implications for cosmological consequences of the same EMSG theory. We draw final conclusions from our results in Sec. \ref{sec:discuss}.

%%%%%%%%%%%%%%%%%%%%%%%%%%%%%%%%%%%%%%%%%%%%%%%%%%
%%%%%%%%%%%%%%%%%%%%%%%%%%%%%%%%%%%%%%%%%%%%%%%%%%

\section{Energy-Momentum Squared Gravity}

\label{sec:EMSG} %%%%%%%%%%%%%%%%%%%%%%%%%%%%%%%%%%%%%%%%%%%%%%%%%%
%%%%%%%%%%%%%%%%%%%%%%%%%%%%%%%%%%%%%%%%%%%%%%%%%%
The EMSG model is constructed by adding a self-contraction of EMT, $T_{\mu \nu }T^{\mu \nu }$, to the Einstein-Hilbert action with a cosmological constant as follows:
\begin{align}
S=\int \left[\frac{1}{2\kappa}\left(\mathcal{R}-2\Lambda\right)+\alpha T_{\mu\nu}T^{\mu\nu}+ \mathcal{L}_{\mathrm{m}}\right]\sqrt{-g}\,\mathrm{d}^4x,
\label{action}
\end{align}
where $\mathcal{R}$ is the scalar curvature, $\kappa =8\pi G$ is the usual gravitational coupling with $G$ being Newton's constant, $\Lambda $ is a cosmological constant, and $\mathcal{L}_{\mathrm{m}}$ is the matter Lagrangian density. The term $T_{\mu \nu }T^{\mu \nu }$ is the EMSG modification with a real constant $\alpha $ that determines the gravitational coupling strength of the modification under consideration.

As usual, we define the EMT as
\begin{align}  \label{tmunudef}
T_{\mu\nu}=-\frac{2}{\sqrt{-g}}\frac{\delta(\sqrt{-g}\mathcal{L}_{\mathrm{m}})}{\delta g^{\mu\nu}}=g_{\mu\nu}\mathcal{L}_{\mathrm{m}}-2\frac{\partial \mathcal{L}_{\mathrm{m}}}{\partial g^{\mu\nu}},
\end{align}
which depends only on the metric tensor components, and not on its derivatives. We consider the perfect fluid form of the EMT given by
\begin{align}  \label{em}
T_{\mu\nu}=(\rho+P)u_{\mu}u_{\nu}+P g_{\mu\nu},
\end{align}
where $\rho $ is the energy density, $P$ is the thermodynamic pressure and $u_{\mu }$ is the four-velocity satisfying the conditions $u_{\mu }u^{\mu }=-1$, $\nabla _{\nu }u^{\mu }u_{\mu }=0$. Unless stated otherwise, we choose units with $\hbar =c=1$ throughout the paper. Varying the action given in Eq. \eqref{action} with respect to the inverse metric, we obtain the modified Einstein's field equations:
\begin{align}
G_{\mu\nu}+\Lambda g_{\mu\nu}=\kappa T_{\mu\nu}+\kappa \alpha \left(g_{\mu\nu}T_{\sigma\epsilon}T^{\sigma\epsilon}-2\theta_{\mu\nu}\right),
\label{fieldeq}
\end{align}
where $G_{\mu \nu }=\mathcal{R}_{\mu \nu }-\frac{1}{2}g_{\mu \nu }\mathcal{R}$ is the Einstein tensor. All contributions from the variation of each new EMT term are collected in the new tensor $\theta _{\mu \nu }$ as
\begin{equation}
\begin{aligned} \theta_{\mu\nu}=& T^{\sigma\epsilon}\frac{\delta
T_{\sigma\epsilon}}{\delta g^{\mu\nu}}+T_{\sigma\epsilon}\frac{\delta
T^{\sigma\epsilon}}{\delta g^{\mu\nu}} \\ =&-2\mathcal{L}_{\rm
m}\left(T_{\mu\nu}-\frac{1}{2}g_{\mu\nu}T\right)-TT_{\mu\nu} \\
&+2T_{\mu}^{\gamma}T_{\nu\gamma}-4T^{\sigma\epsilon}\frac{\partial^2
\mathcal{L}_{\rm m}}{\partial g^{\mu\nu} \partial g^{\sigma\epsilon}}.
\label{theta} \end{aligned}
\end{equation}
Here $T$ is the trace of the EMT. We note that the EMT given in Eq. \eqref{tmunudef} does not include the second variation of $\mathcal{L}_{\mathrm{m}}$, hence the last term of Eq. \eqref{theta} is null. As it is known that the definition of matter Lagrangian giving the perfect fluid EMT is not unique; one could choose either $\mathcal{L}_{\mathrm{m}}=P$ or $\mathcal{L}_{\mathrm{m}}=-\rho $, which provide the same EMT (see \cite{Bertolami:2008ab,Faraoni:2009rk} for a detailed discussion). In the present study, we consider $\mathcal{L}_{\mathrm{m}}=P$. The covariant divergence of Eq. \eqref{fieldeq} reads
\begin{equation}
\nabla^{\mu}T_{\mu\nu}=-\alpha g_{\mu\nu}\nabla^{\mu}
(T_{\sigma\epsilon}T^{\sigma\epsilon})+2\alpha\nabla^{\mu}\theta_{\mu\nu},
\label{nonconservedenergy}
\end{equation}
where we see that local covariant energy-momentum conservation is not satisfied in general, but, is for instance, in the case $\alpha =0$, as it should be (see \cite{ros16,aka17,barrow17} for a further reading).

Substituting Eq. \eqref{em} in Eq. \eqref{theta}, and then using the resultant equation in Eq. \eqref{fieldeq}, we reach the following more illuminating equation
\begin{widetext}
  \begin{align}
G_{\mu\nu}+\Lambda g_{\mu\nu}=\kappa \rho \left[\left(1+\frac{P}{\rho}\right)u_{\mu}u_{\nu}+\frac{P}{\rho}g_{\mu\nu}\right]+\alpha\kappa\rho^2\left[2\left(1+\frac{4P}{\rho}+\frac{3P^2}{\rho^2}\right)u_{\mu}u_{\nu}+\left(1+\frac{3P^2}{\rho^2}\right)g_{\mu\nu}\right].
\label{fieldeq2}
\end{align}
\end{widetext}
We note that the expressions in square brackets on the right-hand side of Eq. \eqref{fieldeq2} are of order unity even for a wide range of realistic sources, namely, e.g. radiation/relativistic matter ($P=\rho /3$) and dust ($P=0$) in between two extremes, Zeldovich (stiff) fluid ($P=\rho$), which is the most rigid EoS compatible with the requirements of relativity theory \cite{zel61} and conventional vacuum energy ($P=-\rho $). This implies that the effect of the new terms due to EMSG modification ($\propto \alpha\kappa\rho ^{2}$) increases almost linearly with respect to the usual terms ($\propto \kappa \rho $) and almost quadratically with respect to $\Lambda $ as values of the energy density $\rho $ increase. This in turn implies that the distinct differences between EMSG and GR would be best observed and thereby constrained, at the highest energy densities. The energy density corresponding to $\Lambda $ is well constrained from cosmological observations as $\rho _{\Lambda }=\Lambda/\kappa \sim 10^{-9}\,\mathrm{erg\,cm^{-3}}$ and is comparable with the energy density scale of the present-day universe $\rho _{\mathrm{cosmic}}\sim 10^{-9}\,\mathrm{erg\,cm^{-3}}$ \cite{Planck15Cosmo}, but is completely negligible, for instance, in comparison with energy density scales of the big bang nucleosynthesis (BBN) $\rho _{\mathrm{bbn}}\sim10^{25}\,\mathrm{erg\,cm^{-3}}$ \cite{Dodelson:1282338} or NSs $\rho _{\mathrm{ns}}\sim 10^{37}\,\mathrm{erg\,cm^{-3}}$ \cite{Shapiro83}. The remaining two terms in brackets in Eq. \eqref{fieldeq2}, on the other hand, are comparable if $\rho \sim\alpha\rho ^{2}$, which implies $|\alpha |\sim\rho ^{-1}$. Hence, the corrections due to the EMSG modification would be observable in the dynamics of the present-day universe if $|\alpha |\sim10^{9}\,\mathrm{erg^{-1}\,cm^{3}}$, affect BBN in the early universe if $|\alpha |\sim 10^{-25}\,\mathrm{erg^{-1}\,cm^{3}}$, and affect compact astrophysical objects like NSs if $|\alpha |\sim 10^{-37}\,\mathrm{erg^{-1}\,cm^{3}}$. Hence, the most stringent constraints on $\alpha $ can be obtained from compact astrophysical objects such as NSs. Black holes, on the other hand, are much denser than NSs, but are not useful for obtaining constraints on EMSG as we mentioned in the Introduction, in Sec. \ref{section:intro}. Thus, in what follows we study constraints on the free parameter $\alpha $ of the EMSG model from NSs and then discuss the implications of the results on the physics of NSs as well as any cosmology based on EMSG.

\section{Hydrostatic equilibrium in EMSG}

\label{sec:hydrostatics}
%%%%%%%%%%%%%%%%%%%%%%%%%%%%%%%%%%%%%%%%%%%%%%%%%%%%%%%%%%%%
%%%%%%%%%%%%%%%%%%%%%%%%%%%%%%%%%%%%%%%%%%%%%%%%%%%%%%%%%%%%55

We seek spherically symmetric solutions of the EMSG field equations inside a nonrotating NS, and so consider a spherically symmetric and static metric in the form
\begin{equation}  \label{eqn:metric}
\mathrm{d} s^2 = -e^{2\nu\left(r\right)}\mathrm{d} t^2 +e^{2\lambda\left(r\right)}\mathrm{d} r^2+r^2\mathrm{d}\theta^2+r^2\sin^2\theta \, \mathrm{d}\phi^2
\end{equation}
with two independent functions $\nu (r)$ and $\lambda \left( r\right) $. Using the metric given in Eq. \eqref{eqn:metric} in Eq. \eqref{fieldeq2}, we reach the following set of field equations, Eqs.\eqref{eqn:f1}-\eqref{eqn:f2},
\begin{equation}
\begin{aligned} \label{eqn:f1}
\frac{1}{r^2}-\frac{e^{-2\lambda}}{r^2}\left(1-2r\frac{\dif \lambda}{\dif r}\right)=\kappa\rho+\kappa\alpha\rho^2\left(1+8\frac{P}{\rho}+3\frac{P^2}{\rho^2}\right),
\end{aligned}
\end{equation}
\begin{align}
-\frac{1}{r^2}+\frac{e^{-2\lambda}}{r^2}\left(1+2r\frac{\dif\nu}{\dif r}\right)=&\kappa P+\kappa\alpha\rho^2\left(1+3\frac{P^2}{\rho^2}\right),  \label{eqn:f2}
\end{align}
where $\rho $ and $P$ are the mass density and pressure at the distance $r$ from the center of NS.

To recast these equations into a more familiar form of the so-called Tolman-Oppenheimer-Volkoff (TOV) equations: first, we define the mass parameter, $m(r),$ within radius $r$ by
\begin{equation}  \label{eqn:pt1}
e^{-2\lambda\left(r\right)}=1-\frac{2Gm\left(r\right)}{r},
\end{equation}
where $m(r)$ is the total mass inside the sphere of radius $r$. The other metric function, $\nu (r)$, is related to the pressure via the radial component of the divergence of the field,
\begin{align}  \label{eqn:pt2}
\frac{\mathrm{d} \nu}{\mathrm{d} r} =&- \left\{\rho \left( 1+\frac{P}{\rho}\right) \left[1+2\alpha\rho\left( 1+3\frac{P}{\rho} \right)\right] \right\}^{-1}  \notag \\
&\times \left[ \left(1+6\alpha P\right) \frac{\mathrm{d} P}{\mathrm{d} r}+2\alpha\rho\frac{\mathrm{d}\rho}{\mathrm{d} r} \right],
\end{align}
with $\nabla ^{\mu }G_{\mu \nu }=0$. Using Eqs. \eqref{eqn:pt1}-\eqref{eqn:pt2}, we find that the modified TOV equations, describing the hydrostatic equilibrium of relativistic stars, now read
\begin{align}
\frac{\mathrm{d} m}{\mathrm{d} r}=4\pi r^2 \rho \left[1+\alpha\rho \left(1+8\frac{P}{\rho}+3\frac{P^2}{\rho^2 }\right)\right],  \label{TOV1}
\end{align}
and

\begin{align}
\frac{\mathrm{d} P}{\mathrm{d} r}&= -G\frac{m\rho }{r^2}\left(1+\frac{P}{\rho}\right) \left( 1-\frac{2Gm}{r}\right)^{-1}  \notag \\
&\times \left[ 1+\frac{4\pi r^3 P}{m }+\alpha \frac{4\pi r^3\rho^2}{m}\left(1+3\frac{P^2}{\rho^2}\right)\right] \notag \\
&\times \left[1+2\alpha\rho\left(1+3\frac{P}{\rho} \right)\right] \left[1 + 2\alpha\rho \left(c_s^{-2}+3\frac{P}{\rho}\right)\right]^{-1},  \label{TOV2}
\end{align}
where $c_{s}^2\equiv\dif P/\dif \rho$ is the sound speed. This set of equations, Eqs. \eqref{TOV1}-\eqref{TOV2}, is closed by an EoS, $P(\rho )$, which prescribes the relation between the pressure $P(r)$ and the density $\rho(r)$.

\section{Method}

\label{sec:method}

We solve the hydrostatic equilibrium equations \eqref{TOV1}-\eqref{TOV2} numerically for a specific EoS, $P=P(\rho )$, by using a $4{\mathrm{th}}$-order Runge-Kutta method \citep{numrec} with radial step-size adapted to the pressure and mass scale-height \citep{bps}. For each EoS, we start with a central density $\rho_{\mathrm{c}}$, and a corresponding central pressure $P_{\mathrm{c}}$, and then integrate towards the surface where the pressure vanishes. This point is defined as the radius of the star, $R$, and $M=m(r=R)$ is the total mass of the star. Then, we change the central density and find the mass and radius of the star again. We repeat the process to determine the mass-radius relation of the star. We then repeat the whole process for different values of $\alpha $ to isolate its effect on the mass-radius relation. We performed all these processes for 4 different representative choices of EoS\footnote{Our choice of the EoS set is representative in the sense that we have employed one sample from each of the large families of EoS: Skyrme models \citep[SkOp;][]{Rei99,Dan09,GR15}, relativistic mean field models \citep[MS2;][]{ref_ms2}, microscopic calculations \citep[APR;][]{ref_APR,bps,ref_SLY} and relativistic mean field models with hyperons \citep[GM1 Y4;][]{ref_GM1,ref_SLY,Oer+15} reflecting the classification of cold NS EoS in CompStar Online Supernov\ae\ Equations of State (\url{https://compose.obspm.fr/eos/48/}).}. These choices reflect the uncertainties that exist in the EoS of NSs. The physical basis of these EoS models, except for SkOp, are discussed in Ref. \cite{lat01}.

\section{Results}

\label{sec:results}

\subsection{Preliminary investigations}

\label{sec:pre}

\begin{figure*}[t!!]
\captionsetup{justification=raggedright,singlelinecheck=false}
\par
\begin{center}
\subfigure[]{    \label{fig:effstiffness}
   \includegraphics[width=0.44\textwidth]{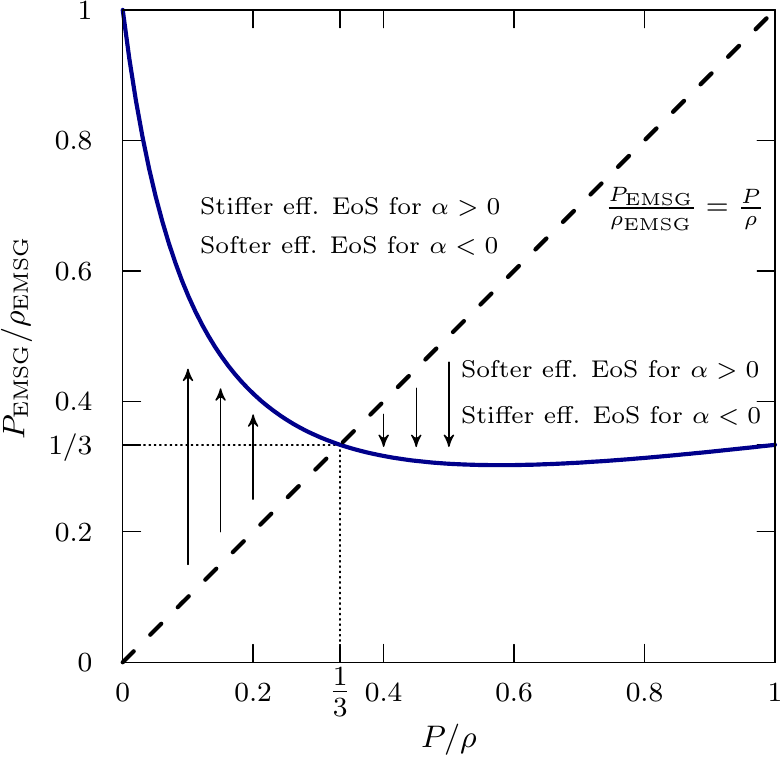}
            }
\subfigure[]{    \label{fig:EoSeff}
   \includegraphics[width=0.43\textwidth]{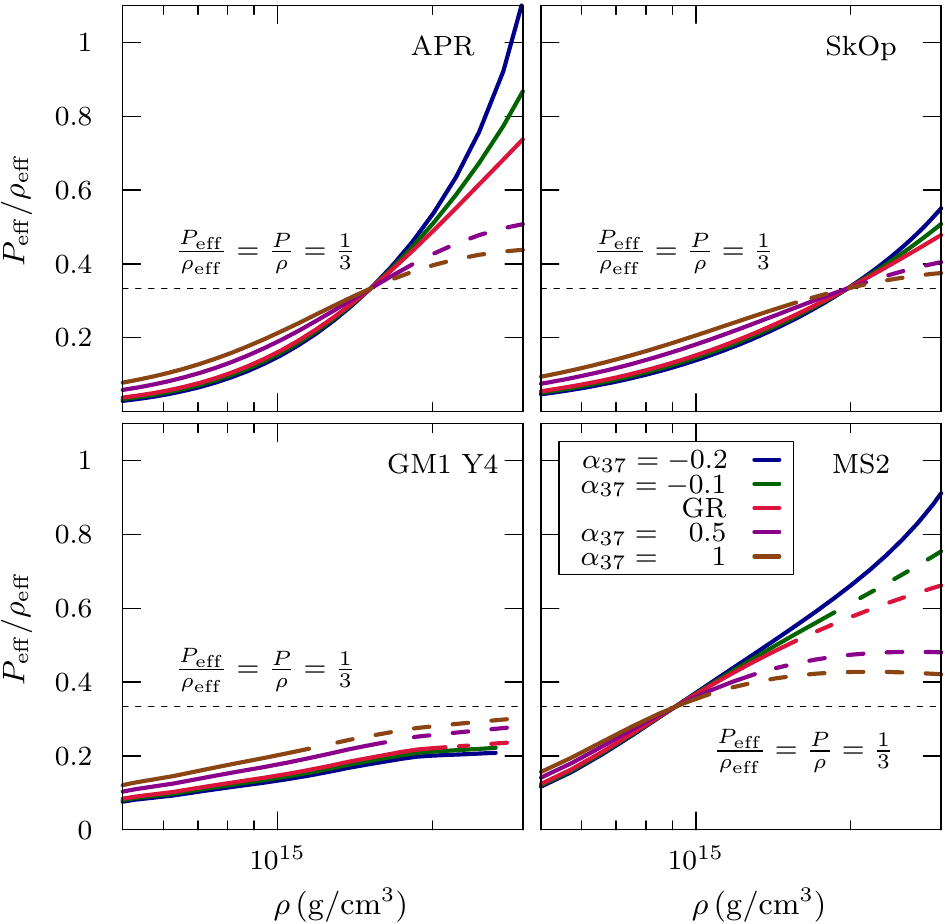}
           }
\end{center}
\caption{ (a) The stiffening/softening of the effective equation of state (EoS), $(P+P_{\mathrm{EMSG}})/(\rho +\rho _{\mathrm{EMSG}})$, due to EMSG with respect to a given EoS. The arrows are added to indicate that $P_{\mathrm{EMSG}}/\rho _{\mathrm{EMSG}}>P/\rho $ for $P/\rho <1/3$ and $P_{\mathrm{EMSG}}/\rho _{\mathrm{EMSG}}<P/\rho $ for $P/\rho >1/3$. The effective stiffness is the same with that of the matter stress at $P/\rho =1/3=P_{\mathrm{EMSG}}/\rho_{\mathrm{EMSG}}$. (b) The effective EoS versus the energy density of the matter stress. The case $\alpha=0$ (GR) gives the EoS of the matter stress itself.}
\label{effective}
\end{figure*}

Before presenting the results of numerical simulations, it is useful to have an estimate of the EMSG modification to GR for typical parameters and possible effects it can play on the structure of NSs. To do so, we define dimensionless modification to GR in Eq. \eqref{TOV1} as
\begin{equation}
\alpha^\prime = \alpha\rho \left(1+8\frac{P}{\rho}+3\frac{P^2}{\rho^2 }\right).  \label{alphap}
\end{equation}
We investigate the contribution of this term to the terms in brackets in Eq. \eqref{TOV1} for a typical NS of mass $M=1.5M_{\odot }$ and radius $R=10^{6}\,\mathrm{cm}$ whose central density is $\sim 10^{16}\,\mathrm{g\,cm^{-3}}$ ($\rho \sim 10^{37}\,\mathrm{erg\,cm^{-3}}$). The maximum value of $P/\rho $, attained at the center of a typical NS, is about $0.2$ \cite{eks16} so that the value of the term in parenthesis in Eq. \eqref{alphap} is about $2.7$. We see from this analysis that the absolute value of $\alpha$ should be less than $10^{-37}\,\mathrm{erg^{-1}\,cm^{3}}$, otherwise $\alpha^{\prime }$ is of order unity and creates strong deviations from the predictions of GR. We would expect perturbative modifications to the structure of the star for values of $\alpha ^{\prime }$ one or two orders of magnitude less than this nominal value. For values of $\alpha ^{\prime}$ even less than this value, the structure of NS within the theory is not expected to differ significantly from what one obtains within GR.

The cumulative mass $m(r)$ should increase monotonically with $r$ as spherical shells of matter are added in the integration process. This is guaranteed in Newtonian gravity and GR, where $\mathrm{d}m/\mathrm{d}r=4\pi r^{2}\rho $ i.e.\ the right-hand side is positive definite. This would not be satisfied in the EMSG model of gravity if $\alpha^{\prime }<-1$ (see Eq. \eqref{TOV1} and Eq. \eqref{alphap} together). We note that because we expect the value of the EoS parameter, $P/\rho$, to reach its highest value at the center of the star and both the EoS parameter and density decrease as we move away from the center to the surface of the star, once this  condition is satisfied at the center of the star then it is guaranteed that it would not be violated anywhere else within the star. Hence, one should impose the following condition to alleviate the ``negative mass shell'' problem
\begin{equation}
\alpha > -\frac{1}{\rho_{\mathrm{c}} \left(1+8\frac{P_{\mathrm{c}}}{\rho_{ \mathrm{c}}}+3\frac{P_{\mathrm{c}}^2}{\rho_{\mathrm{c}}^2 }\right)} \quad \mbox{for $\mathrm{d}m/\mathrm{d}r>0$}.  \label{crit1}
\end{equation}
This implies, given $P_{\mathrm{c}}/\rho _{\mathrm{c}}\sim 0.2$ and $\rho _{ \mathrm{c}}\sim 10^{37}\,\mathrm{erg\,cm^{-3}}$ at the center of a typical NS, that $\alpha \gtrsim -0.38\times 10^{-37}\,\mathrm{erg^{-1}\,cm^{3}}$; otherwise the model would contradict with the existence of relativistic
stars.

In addition, in order that the star is stable, the pressure should decrease monotonically outwards ($\mathrm{d}P/\mathrm{d}r<0$) with the radial coordinate $r$. This kind of stability is again guaranteed in Newtonian gravity where $\mathrm{d}P/\mathrm{d}r=-Gm\rho /r^{2},$ and also in GR$-$where all relativistic correction terms are positive (see Eq. \eqref{TOV2} with $\alpha =0$). The presence of $\alpha$ terms in Eq. \eqref{TOV2} risks the stable stratification of the star, particularly, when it is allowed to take negative values. Thus, to avoid such an issue in EMSG, noticing in Eq. \eqref{TOV2} that the last term with square brackets reaches negative values before the other multipliers as $\alpha$ is given larger negative values, we expect
\begin{equation}
\alpha > -\frac{1}{6P+2\rho\, c_s^{-2}} \qquad \mbox{for $\mathrm{d}P/\mathrm{d}r<0$,}
\label{crit2}
\end{equation}
which leads to $\alpha \gtrsim -10^{-38}\,\mathrm{erg^{-1}\,cm^{3}}$ when we consider central values of the parameters for a typical NS as done above. We note here that these values are estimates using some typical values for NSs, the precise results will be obtained numerically below.

Modifications to the hydrostatic equilibrium due to EMSG  can reveal some features of the influence of EMSG on the NSs configurations by considering them together with the usual terms that appear in GR. We see from Eqs. \eqref{eqn:f1}-\eqref{eqn:f2} that the additional energy density and pressure terms, $\rho _{\mathrm{EMSG}}=\alpha (\rho ^{2}+8\rho P+3P^{2})$ and $P_{\mathrm{EMSG}}=\alpha (\rho ^{2}+3P^{2})$, respectively, arising from EMSG yield an EoS parameter,
\begin{equation}
\frac{P_{\mathrm{EMSG}}}{\rho _{\mathrm{EMSG}}}=1-\left[ \frac{1}{8}\frac{\rho }{P}+1+\frac{3}{8}\frac{P}{\rho }\right] ^{-1},
\end{equation}
with a range of $P_{\mathrm{EMSG}}/\rho _{\mathrm{EMSG}}=[1,1/3]$ for $P/\rho =[0,1/3]$ and gives $P_{\mathrm{EMSG}}/\rho _{\mathrm{EMSG}}\lesssim 1/3$ for $P/\rho =[1/3,1]$. Therefore, $P_{\mathrm{EMSG}}/\rho _{\mathrm{EMSG}}>P/\rho $ for $P/\rho <1/3$, $P_{\mathrm{EMSG}}/\rho _{\mathrm{EMSG}}=P/\rho $ for $P/\rho =1/3$ (viz., a critical point), and $P_{\mathrm{EMSG}}/\rho _{\mathrm{EMSG}}<P/\rho $ for $P/\rho >1/3$. This implies that, within a NS, these new terms stiffen the effective EoS parameter $P_{\rm eff}/\rho_{\rm eff}$ [where $P_{\rm eff}=P+P_{ \mathrm{EMSG}}$ and $\rho_{\rm eff}=\rho +\rho _{\mathrm{EMSG}}$] when $P/\rho <1/3$, and soften it when $P/\rho <1/3$ if $\alpha >0$, and conversely soften the effective EoS when $P/\rho <1/3$ and stiffen it when $P/\rho >1/3$ if $\alpha <0$. We show in Fig.~\ref{fig:effstiffness}, and indicate using arrows, the effective stiffening/softening due to EMSG for the range $P/\rho=[0,1]$ depending on the sign of $\alpha $. Note that $P/\rho =1/3$ is the critical point where the effective EoS is the same as the EoS of the matter stress. It is conceivable that the effective stiffening may compensate the softening of the matter stress due to the hyperonization and so may enhance the maximum mass of NSs with hyperons. The situation in general, however, is more complicated for an EoS parameterization that can reach values higher than $1/3$. Given that the matter stresses tend to become stiffer with depth in a NS, EMSG for $\alpha >0$ would lead to effective stiffening down to a certain depth (before which $P/\rho<1/3$) and thereafter ($P/\rho>1/3$) to effective softening at further depths (and vice versa for $\alpha <0$). Thus, how EMSG would modify NS configurations overall, and whether it can address the hyperon puzzle or not, should be investigated by full numerical solutions of NSs in EMSG for various realistic EoS parametrizations, which will also allow us to constrain the free parameter $\alpha $ of EMSG.

%%%%%%%%%%%%%%%%%%%%%%%%%%%%%%%%%%%%%%%%%%%%%%%%%%%%%%%%%%%%

\begin{figure}[t!!]
\captionsetup{justification=raggedright,singlelinecheck=false}
\centering
\includegraphics[width=0.43\textwidth]{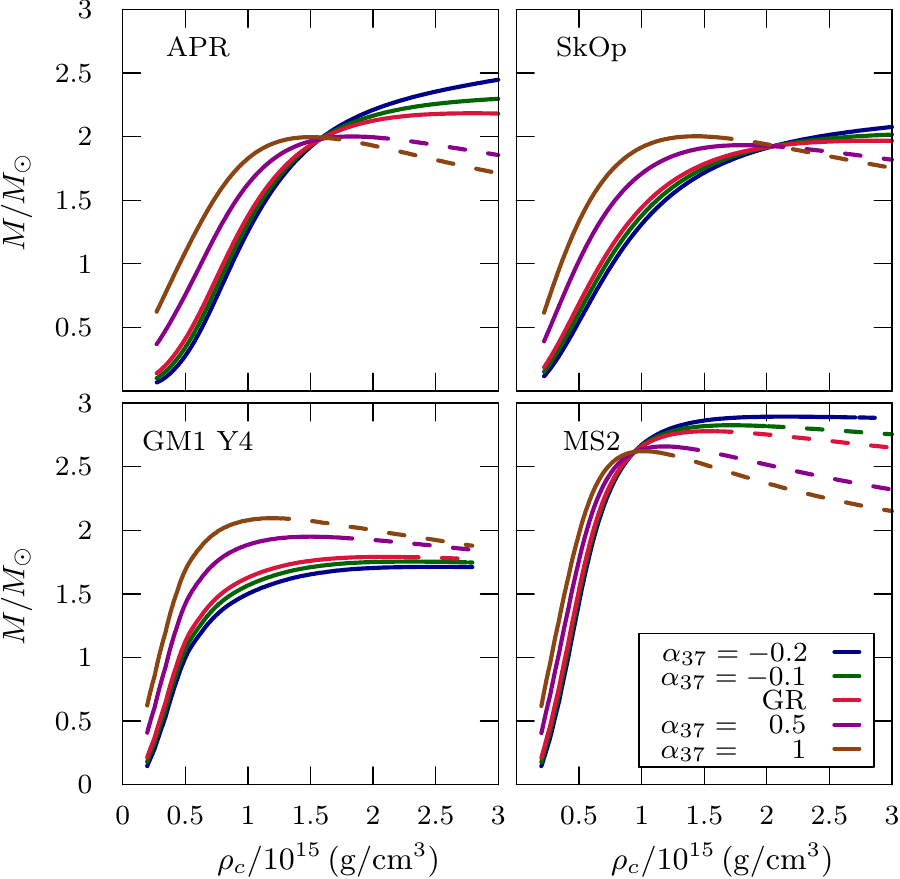}  \caption{Mass-central density ($M$-$\rho_{\rm c}$) relations for various EoS' for a range of $\alpha$ values. The solid lines correspond to stable, the dashed lines to unstable solutions that do not satisfy the stability criteria given in Eq. \eqref{stability_star}.}
\label{fig:Mrhoc}
\end{figure}

\subsection{The mass-radius relations}

%%%%%%%%%%%%%%%%%%%%%%%%%%%%%%%%%%%%%%%%%%%%%%%%%%%%%%%%%%%%

The compactness and curvature parameters within a NS are orders of magnitude larger than their values in the solar system \cite{eks+14}. This allows us to constrain free parameters of some modified models of gravity by using mass-radius measurements of NSs (see e.g.\ \cite{ara+11,del+12}). Yet one cannot put precise limits on the free parameters of modified gravity models$-$not only because the simultaneous mass and radius measurements are not yet precise$-$because the EoS prevailing at the core of NSs is not well constrained by nuclear collision experiments. The central density is an order of magnitude larger than that probed in heavy-ion collision experiments and the EoS is very sensitive to the nuclear symmetry energy and its slope at the saturation density \cite{he+15}. The two observables, mass and radius, of NSs are determined by both the model of gravity and EoS leading to the so called ``degeneracy'' issue hindering high precision constraints on models of gravity.

\begin{figure*}[t!!]
\captionsetup{justification=raggedright,singlelinecheck=false}
\includegraphics[width=0.65\textwidth]{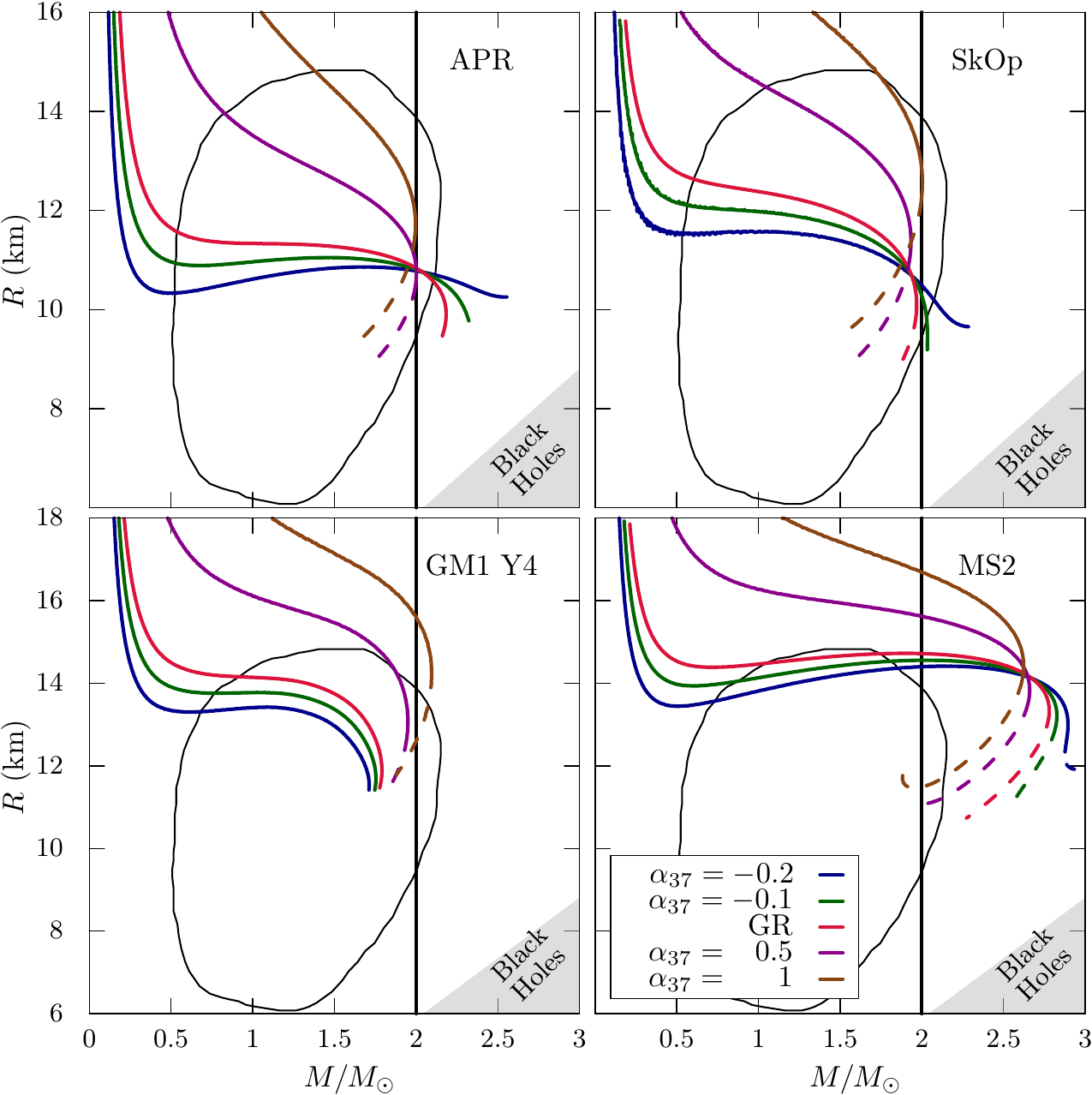}
\caption{Mass-radius ($M$-$R$) relations for various EoS'. Here, $\alpha_{37}$ equals to $\alpha/(10^{-37}\,\mathrm{cm^3/erg})$. The elliptical curve corresponds to the combined constraints (68\%) obtained by mass radius measurements of NSs in low-mass x-ray binaries (see Fig. \ref{fig:constraint} in \cite{oze16} and references therein). The thick solid black line is the highest precisely measured mass, $M\simeq 2\,M_{\odot}$ of a NS \citep{dem+10,ant+13}.}
\label{fig:MR}
\end{figure*}

We follow the procedure presented in Sec. \ref{sec:method} to obtain the mass-central density ($M$-$\rho_{\mathrm{c}}$) (see Fig. \ref{fig:Mrhoc}) and mass-radius ($M$-$R$) relations (see Fig.\ref{fig:MR}) of NSs within the framework of the EMSG gravity model.

\begin{figure}[h!!]
\captionsetup{justification=raggedright,singlelinecheck=false}
\includegraphics[width=0.43\textwidth]{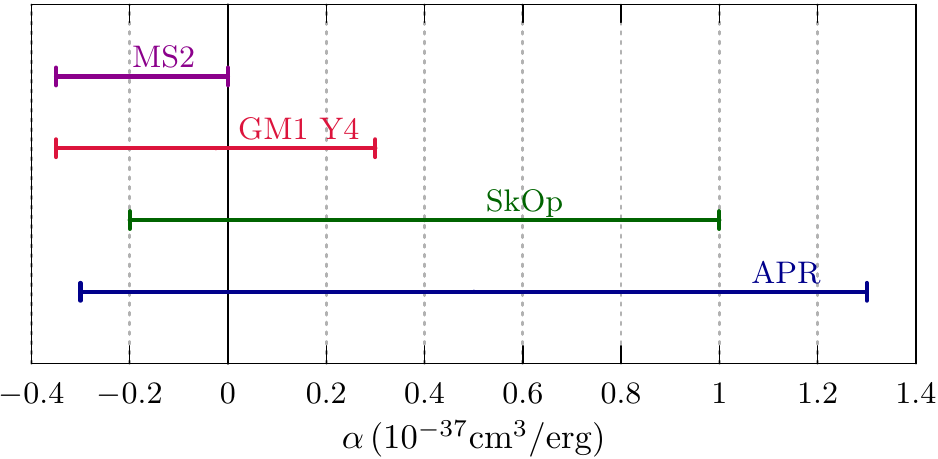}
\caption{The range of $\alpha$ consistent with the observations for each EoS studied.} \label{fig:constraint}
\end{figure}

In order to be viable, a mass-radius relation has to pass through the elliptical curve corresponding to the combined constraints (at 68\% confidence level) obtained by mass-radius measurements of NSs in low-mass x-ray binaries (see Fig. \ref{fig:constraint} in \cite{oze16} and references therein) as well as attain a maximum mass exceeding two solar-masses since NSs with such masses are observed to exist \citep{dem+10,ant+13}.

Given that the EoS of the NS is not strongly constrained by terrestrial experiments, we employ four different representative choices of EoS---APR \citep{ref_APR,bps,ref_SLY}, SkOp \citep{Rei99,Dan09,GR15}, GM1 Y4 \citep{ref_GM1,ref_SLY,Oer+15} and MS2 \citep{ref_ms2} ---to isolate the implications of the gravity model. The results are summarized as follows:

\begin{enumerate}[i)]
\item APR:  APR is an EoS derived by variational techniques and it assumes the presence of only hadronic matter (no hyperons). The stiffness of this EoS can pass above $1/3$ when the density exceeds a certain value, that allows this parameterization to give $2.18\,M_{\odot }$ for the maximum mass of a NS within GR. The effective stiffening/softening throughout the NS due to EMSG, which would change depending on the matter stress EoS varying as its energy density changes with the depth within the NS (see Fig. \ref{effective}), consequently leads the maximum mass of the NSs to increase/decrease for negative/positive values of $\alpha $. When $\alpha \lesssim -0.3\times 10^{-37}\,\mathrm{erg^{-1}\,cm^{3}}$, no NS solution can be obtained with APR as explained in Sec. \ref{sec:pre}. The maximum mass has a minimum, $1.99\,M_{\odot }$, at $\alpha \simeq 0.8\times 10^{-37}\,\mathrm{erg^{-1}\,cm^{3}}$. Greater values of $\alpha $ increase the maximum mass, but the solutions then cannot satisfy the $M$-$R$ constraint taken from Ref.~\cite{oze16} for $\alpha \gtrsim 1.3\times 10^{-37}\,\mathrm{erg^{-1}\,cm^{3}}$.

\item SkOp: SkOp is an EoS taking into account the Skyrme interactions in the presence of only hadronic matter (no hyperons). The stiffness of this EoS can pass slightly above $1/3$ when the density exceeds a certain value, that leads this parameterization to give $1.96\,M_{\odot }$ for the maximum mass of a NS within GR. The effective stiffening/softening throughout the NS due to EMSG, which would change depending on the matter stress' EoS varying as its energy density changes with the depth within the NS (see Fig. \ref{effective}), consequently leads the maximum mass of the NSs to increase/decrease for negative/positive values of $\alpha $, and leads it to exceed $2\,M_{\odot }$ when $\alpha \lesssim -0.06\times 10^{-37}\,\mathrm{erg^{-1}\,cm^{3}}$. When $\alpha \lesssim -0.2\times 10^{-37}\,\mathrm{erg^{-1}\,cm^{3}}$, no NS solution can be obtained with SkOp, as explained in Sec. \ref{sec:pre}. The maximum mass has a minimum, $1.92\,M_{\odot }$, at $\alpha \simeq 0.3\times 10^{-37}\,\mathrm{erg^{-1}\,cm^{3}}$. The solutions cannot satisfy the $M$-$R $ constraint taken from Ref.~\cite{oze16} for $\alpha \gtrsim 1\times 10^{-37}\,\mathrm{erg^{-1}\,cm^{3}}$.

\item GM1 Y4: GM1 Y4 is an EoS obtained in the framework of relativistic mean field theory. It allows for the appearance of hyperons along with the presence of hadronic matter. The stiffness of this EoS can never pass above $1/3$, that leads this parametrization to give only $1.79\,M_{\odot }$ for the maximum mass of a NS within GR leading to the so called \textit{hyperon puzzle}. In contrast to the case with the other three EoS parametrizations of matter stress (because in this case the matter stress EoS is always less than $1/3$ throughout the NS) EMSG leads either only to effective stiffening (in case $\alpha >0$) or only to effective softening (in case $\alpha <0$) throughout the NS (see Fig. \ref{effective}). Consequently, EMSG leads to the maximum mass of NSs monotonically increasing with increasing $\alpha $ values. The maximum mass exceeds $2\,M_{\odot }$ at $\alpha \simeq 0.68\times 10^{-37}\,\mathrm{erg^{-1}\,cm^{3},}$ so providing a possible resolution to the \textit{hyperon puzzle} at the expense of very large NS radii. As such, these solutions cannot satisfy the $M$-$R$ constraint given in Ref.~\cite{oze16} for $\alpha \gtrsim 3\times 10^{-37}\,\mathrm{erg^{-1}\,cm^{3}}$. No NS solution can be obtained with GM1 Y4 for $\alpha \lesssim -0.35\times10^{-37}\,\mathrm{erg^{-1}\,cm^{3}}$, as explained in Sec. \ref{sec:pre}.

\item MS2: MS2 is an EoS derived through relativistic mean-field theory considering the presence of only hadronic matter (no hyperons). The stiffness of this EoS can pass above $1/3$ when the density exceeds a certain value, that leads this parametrization to give $2.78\,M_{\odot }$ for the maximum mass of a NS within GR. The effective stiffening/softening throughout the NS due to EMSG, which would change depending on the matter stress EoS varying as its energy density changes with the depth within the NS (see Fig. \ref{effective}), consequently leads the maximum mass of the NSs to increase/decrease for negative/positive values of $\alpha$. The maximum mass does not have a minimum for $\alpha \lesssim 1.3\times10^{-37}\,\mathrm{erg^{-1}\,cm^{3}}$ and decreases with increasing $\alpha $. Because of the quite large radii it predicts, this EoS within GR can only marginally satisfy the $M$-$R$ constraint given in Ref.~\cite{oze16}. On the other hand, EMSG, for the negative values of $\alpha $, leads to NSs with smaller radii resulting with better match to the $M$-$R$ constraints given in Ref.~\cite{oze16}. For $\alpha \lesssim-0.35\times 10^{-37}\,\mathrm{erg^{-1}\,cm^{3}}$, no NS solution can be obtained with MS2 as explained in Sec. \ref{sec:pre}.

\end{enumerate}

%%%%%%%%%%%%%%%%%%%%%%%%%%%%%%%%%%%%%%%%%%%%%%

In summary, our results show that for some values of $\alpha$, compatibility of SkOp and MS2 with the $M$-$R$ observations gets better compared to GR. APR, within GR, is already consistent with observations and a constraint on $\alpha $ from the mass-radius observations and maximum observed mass is
\begin{equation}
-10^{-38}\,\mathrm{cm^{3}/erg}\lesssim \alpha \lesssim 10^{-37}\,\mathrm{cm^{3}/erg}  \label{constraint1}
\end{equation}
as shown in Fig. \ref{fig:constraint}. The hyperonic EoS GM1 Y4 is discussed in the following separately.

%%%%%%%%%%%%%%%%%%%%%%%%%%%%%%%%%%%%%%%%%%%%%

\subsubsection{Stability of the solutions}
\label{sec:stability}

Apart from the local stability of the hydrostatic equilibrium of a mass distribution, given by the conditions $\dif m/\dif r>0$ and $\dif P/\dif r<0$, which is required to be satisfied at every point of a star, we also consider the so called \textit{static stability criterion}  \cite{har+65}
\begin{equation}
\frac{\dif M}{\dif \rho_{\rm c}}>0,  \label{stability_star}
\end{equation}
to be satisfied by all stellar configurations. This is a necessary but not sufficient condition for stability.
Yet a solution satisfying this criterion is unstable only if the solution passes from a critical point (an extremum) on the $M(R)$ curve.
The solution branch we consider in this work coincides with the GR solution at low densities at which the differences between EMSG and GR vanish.
These solutions are then stable up to the point where the above condition, equivalent to $\dif M / \dif R < 0$, is no longer satisfied.

\begin{figure}[h!!]
\captionsetup{justification=raggedright,singlelinecheck=false}
\includegraphics[width=0.43\textwidth]{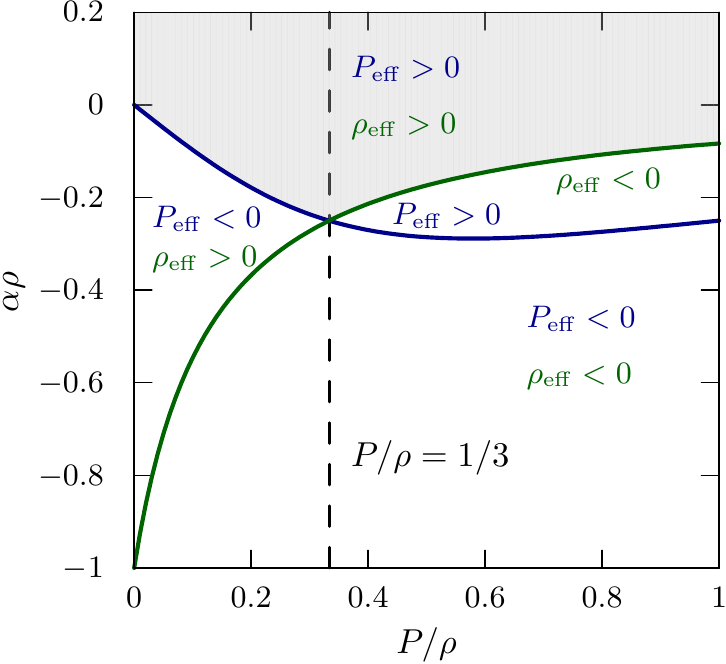}
\caption{ Parameter space allowing for employing the stability criteria including Eq. \eqref{stability_star}. The shaded region shows the parameter space satisfying the criteria given in Eqs. \eqref{sercans_criterion1} and  \eqref{sercans_criterion2}. Here, the green and the blue lines are boundaries where $\rho_{\rm eff}=\rho+\rho_{\rm EMSG}$ and $P_{\rm eff}=P+P_{\rm EMSG}$ change sign, respectively. }
\label{fig:DynStab}
\end{figure}

We present a detailed analysis for the stability of stellar configurations within this model of gravity  in Appendix \ref{sec:apndx0}.
Our results imply that in order to use the stability criteria we mentioned in the previous paragraph within the framework of EMSG the following conditions should be satisfied:
\begin{align}
P+P_{\rm EMSG}>0 \quad \Leftrightarrow\quad \alpha \rho &> - \frac{P/\rho}{1+3 P^2/\rho^2},  \label{sercans_criterion1}  \\
\rho+\rho_{\rm EMSG}>0 \quad\Leftrightarrow\quad \alpha \rho &> -\frac{1}{1+8P/\rho+3P^2/\rho^2}.\label{sercans_criterion2}
\end{align}
We note that these conditions are trivially satisfied for the case $\alpha>0$. The case $\alpha<0$ on the other hand should be investigated carefully: The second  condition $\rho+\rho_{\rm EMSG}>0$ (which is ensured by $\dif m/\dif r>0$) guarantees that $P+P_{\rm EMSG}>0$ for $P/\rho>1/3$ as shown in Fig. \ref{fig:DynStab}. Yet, for $P/\rho<1/3$ the stability of the configurations is not guaranteed. So, we employed the condition $P+P_{\rm EMSG}>0$ in our code and found that it is satisfied in all of our solutions, as shown in Fig. \ref{fig:EoSeff}. Also, $P+P_{\rm EMSG}>0$ is satisfied for smaller densities at which EMSG reduces to GR  [such low densities are not presented in Fig. \ref{fig:EoSeff}].
The solid lines in Figs. \ref{fig:Mrhoc} and \ref{fig:MR} correspond to stable configurations for which the stability criteria including Eq. \eqref{stability_star} are satisfied while the dashed lines correspond to the unstable solutions.

%%%%%%%%%%%%%%%%%%%%%%%%%%%%%%%%%%%%%%%%%%%%%
%%%%%%%%%%%%%%%%%%%%%%%%%%%%%%%%%%%%%%%%%%%%%%%%%%%%%%%%%%%%

%%%%%%%%%%%%%%%%%%%%%%%%%%%%%%%%%%%%%%%%%%%%%%%%%%%%%%%%
\begin{figure*}[t!!]
\captionsetup{justification=raggedright,singlelinecheck=false}
\par
\begin{center}
\subfigure[]{            \label{Mmax}
            \includegraphics[width=0.3\textwidth]{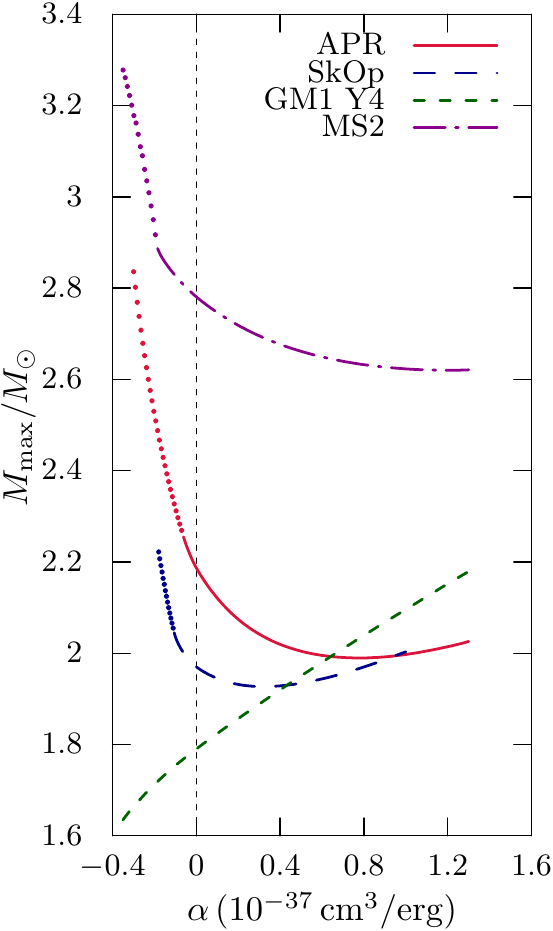}
        }
\subfigure[]{           \label{Rmin}
           \includegraphics[width=0.3\textwidth]{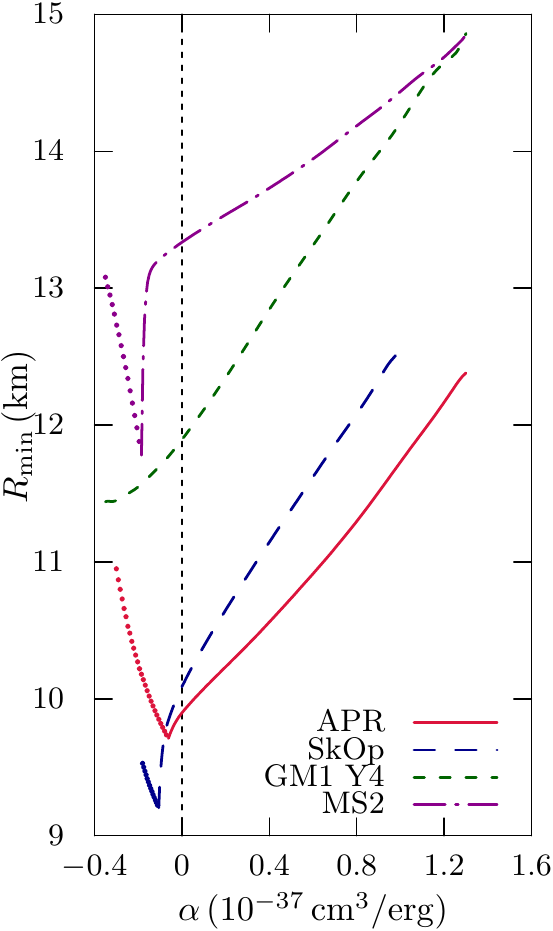}
        }
\subfigure[]{           \label{Comp}
           \includegraphics[width=0.3\textwidth]{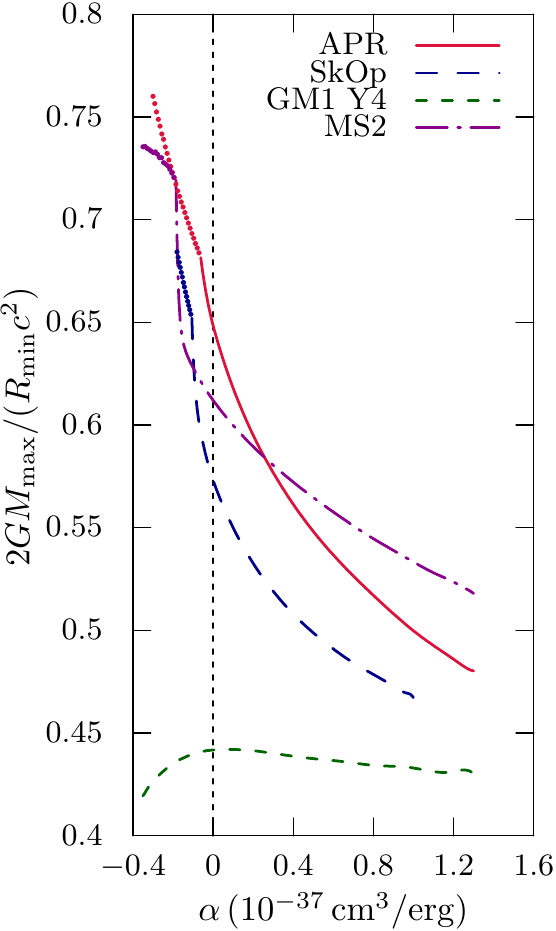}
        }
\end{center}
\caption{(a) Maximum mass of the star at which $\mathrm{d} M/\mathrm{d} \rho_c=0$ vs.\ $\alpha$. (b) The radius of the maximum mass stellar configuration vs.\ $\alpha$. (c) Compactness of the maximum mass stellar configuration vs.\ $\alpha$. The lower bound of $\alpha$ arises from the condition given in Eq. \eqref{crit1} demanding that $\mathrm{d} m/\mathrm{d} r>0$ within the star. Dots correspond to values of $\alpha$ beyond which the maximum mass of the NS is no longer an extremum condition. }
\label{fig:Mmax}
\end{figure*}
%%%%%%%%%%%%%%%%%%%%%%%%%%%%%%%%%%%%%%%%%%%%%%%%%%%%%%%%%%%%%

\subsection{Maximum mass and the hyperon puzzle}

%%%%%%%%%%%%%%%%%%%%%%%%%%%%%%%%%%%%%%%%%%%%%%%%%%%%%%%%%%%%

In GR, pressure not only balances the self-gravity of the star but also acts as a source of gravity. The consequence of this is the existence of a maximum mass for NSs beyond which further increase in pressure destabilizes the star rather than balancing gravity as demonstrated in Fig. \ref{fig:Mrhoc}. Similar to the situation in GR, NSs in the EMSG gravity model under consideration attain a maximum mass, depending on the value of $\alpha $, beyond which the solutions are unstable with respect to the criterion given in Eq. \eqref{stability_star}.

We have studied the dependence of the maximum mass of NSs on the value of $\alpha$, as shown in Fig. \ref{Mmax}. The star attains its minimum radius at this critical mass. We show the dependence of the minimum radius on the value of $\alpha$ in Fig. \ref{Rmin}. The compactness of the star, $\eta \equiv 2GM/R$, at the maximum mass/minimum radius is shown in Fig. \ref{Comp}. We note that the lower bounds of $\alpha$ in Fig. \ref{fig:Mmax} arise from the condition given in Eq. \eqref{crit1} which guarantees that $\mathrm{d} m/\mathrm{d} r>0$ within the star.

These results show that GR ($\alpha =0$) is not special among the family of stellar solutions parametrized by $\alpha $ in this gravity model. We have seen in Fig. \ref{fig:Mmax} that the influence of EMSG on the maximum mass, the minimum radius and the maximum compactness is not trivial. This is the case in particular for APR, SkOp and MS2 parametrizations for which the EoS parameter can pass above the critical value $1/3$ at a certain depth of the NS and therefore the maximum mass (and the minimum radius) would be determined by the interplay of the effective stiffening and softening due to EMSG at different radial coordinate $r$. However, overall we see that the NSs achieve higher masses for negative $\alpha $ values for these EoS'. On the other hand, in case of the hyperonic EoS GM1 Y4 parametrization, we see that maximum mass increases simply monotonically with increasing $\alpha$ values, since this EoS always remains below the critical EoS parameter value $1/3$ and hence EMSG renders the EoS effectively stiffer/softer for positive/negative $\alpha$ values all the way down from the surface to the center.

We see that the maximum mass within SkOp can exceed $2\,M_{\odot }$ limit for some negative values of $\alpha $. In this case the maximum masses are not extremal values and we note that there is no analogue of this solution branch in GR. We have seen that, in case of GM1 Y4 allowing the appearance of hyperons, the maximum mass increases, with respect to its value in GR, for the positive values of $\alpha$ and can exceed $2\,M_{\odot}$ at sufficiently large positive values of $\alpha$. The question, thus, naturally arises whether it is possible to resolve the \textit{hyperon puzzle} within the framework of EMSG. Our results, however, show that the increase in mass with increasing $\alpha $ values does not allow for a satisfactory resolution of the \textit{hyperon puzzle} as it predicts very large radii which are incompatible with the observed radii in Ref.~\cite{oze16}.

\section{Cosmological Implications}
\label{cosmo_analysis}
In this section, we outline the cosmological implications of the EMSG under the constraints Eq. \eqref{constraint1} we obtained for the free parameter of the model $\alpha $ from NSs. We discussed above that the EMSG modification becomes more influential with the increasing energy density values so that it is conceivable that we would see the effect of the EMSG modification in the early stages of the universe. In the early universe we can assume that radiation (photons, gravitons, relativistic massive particles) is dominant, and the spatial curvature and cosmological constant are negligible. Hence, from Eq. \eqref{fieldeq2}, we obtain the cosmological field equations for the EoS $P=\rho /3$ and $\Lambda =0$ within the metric framework of the homogeneous and isotropic spacetime with Euclidean spacelike sections, $\mathrm{d}s^{2}=-\mathrm{d}t^{2}+a(t)^{2}\,\mathrm{d}\vec{x}^{2}$, where $a(t)$ is the cosmic scale factor, as follows:
\begin{equation}
\begin{aligned}
\label{eq:rhopresprime}
3H^2=&\,\,\kappa\rho_{\rm r}+4\alpha\,\kappa\rho_{\rm r}^2\,\,=\kappa\rho_{\rm r}(1+4\alpha\rho_{\rm r}),\\
-2\dot{H}-3H^2=&\,\kappa \frac{\rho_{\rm r}}{3}+\frac{4}{3}\alpha\,\kappa \rho_{\rm r}^2=\kappa \frac{\rho_{\rm r}}{3}(1+4\alpha\rho_{\rm r}),
\end{aligned}
\end{equation}
where $H=\dot{a}/a$ is the Hubble parameter. A realistic\footnote{One may see Ref. \cite{barrow17} for a comprehensive analysis of realistic cosmological solutions in the EMSG. However, for the completeness of our discussion here, we should mention the other solution of the system Eq. \eqref{eq:rhopresprime} as $a=a_{1}\,t^{\frac{1}{2}}$ and $\rho _{\mathrm{r}}=-\frac{1}{8\alpha }\left( 1+\sqrt{1+\frac{12\alpha }{\kappa t^{2}}}\right) $. We note however that, in this case, $\rho _{\mathrm{r}}>0$ only if $\alpha >0$ and more importantly $\lim_{\alpha \rightarrow 0^{+}}\rho _{\mathrm{r}}=\infty $ rather than approaching $\frac{3}{4\kappa t^{2}}$ that would occur in GR. This implies that from this solution we are not able to recover GR completely.} solution of this system of equations Eq. \eqref{eq:rhopresprime} reads:
\begin{equation}
a=a_{1}\,t^{\frac{1}{2}}\quad \textnormal{and}\quad \rho _{\mathrm{r}}=\frac{1}{8\alpha }\left( \sqrt{1+\frac{12\alpha }{\kappa t^{2}}}-1\right) ,
\label{bcksol}
\end{equation}
where $a_{1}$ is the length of the cosmic scale factor when the age of the universe is $t=1\,\mathrm{s}$. We showed in Appendix \ref{sec:apndx} that this solution is stable against linear perturbations.

\begin{figure}[ht!]
\captionsetup{justification=raggedright,singlelinecheck=false}
\par
\begin{center}
\subfigure[]{            \label{rho_h}
            \includegraphics[width=0.43\textwidth]{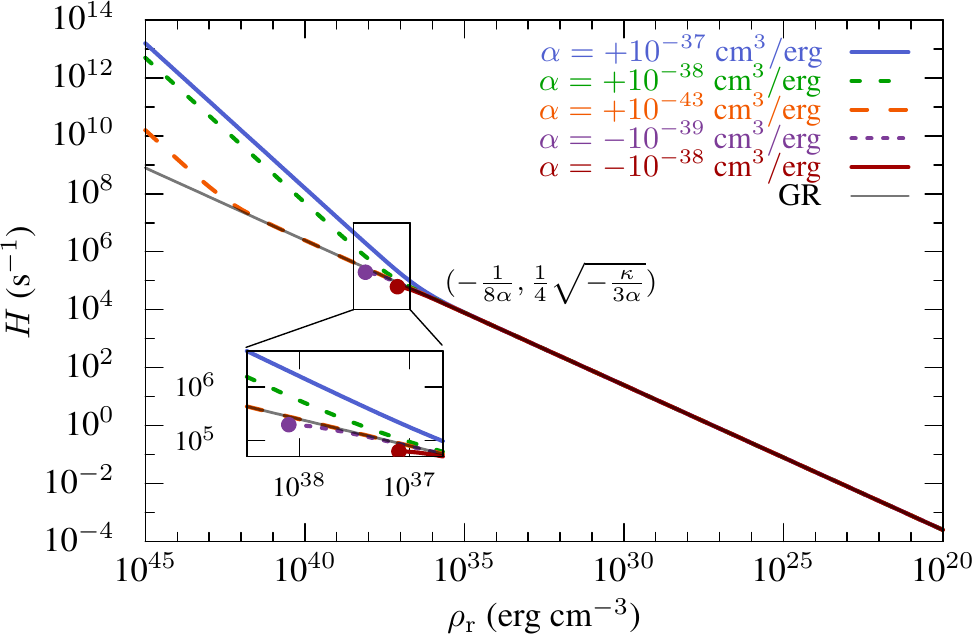}
        }
\subfigure[]{           \label{rho_r}
           \includegraphics[width=0.43\textwidth]{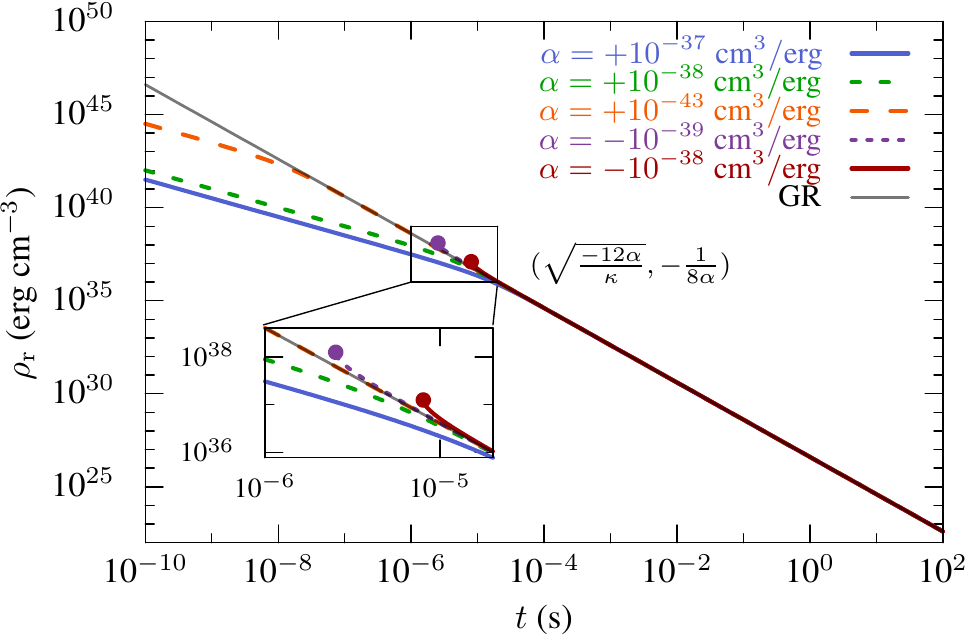}
        }
\subfigure[]{            \label{devfromGRr}
            \includegraphics[width=0.43\textwidth]{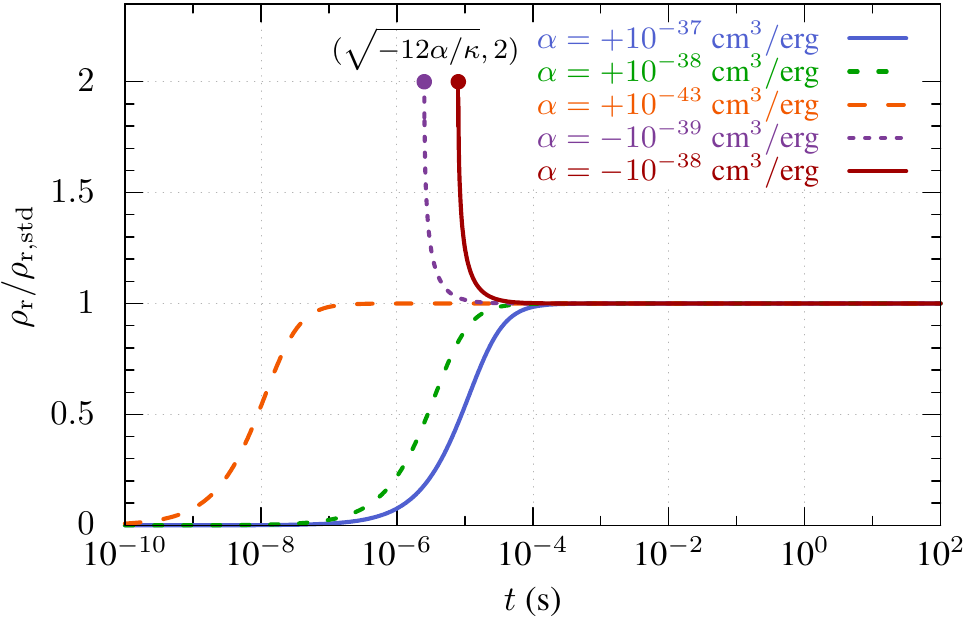}
        }
\end{center}
\caption{ (a) Hubble parameter $H$ versus radiation energy density $\rho_{\mathrm{r}}$. (b) Radiation energy density $\rho_{\mathrm{r}}$ versus cosmic time $t$. (c) The ratio of the modified radiation energy density with respect to the radiation energy density in standard GR. For the limit $\alpha=-10^{-38}\,\mathrm{cm^3/erg}$ there is a nonsingular beginning at $t\sim10^{-4}\,\mathrm{s}$. }
\label{fig:hubble}
\end{figure}

We note that it is the time evolution of the energy density that is modified with respect to the standard radiation dominated universe based on GR giving  $\rho _{\mathrm{r}}=\frac{3}{4\kappa t^{2}}$. One may check that, in the limit $\alpha \rightarrow 0$ in our solution $\rho _{\mathrm{r}}\rightarrow\frac{3}{4\kappa t^{2}}$ and $\rho _{\mathrm{r}}\rightarrow \frac{3a_{1}^{4}}{4\kappa }a^{-4}$, we recover the standard radiation dominated universe. And, for non-zero values of $\alpha $ we have the following two cases: (a) If $\alpha >0$, then $a\rightarrow +\infty $ and $\rho _{\mathrm{r}}\rightarrow 0$ as $t\rightarrow +\infty $ with $a\rightarrow 0$ and $\rho _{\mathrm{r}}\rightarrow +\infty $ as $t\rightarrow 0$ ( ``big bang"). (b) If $\alpha <0$, then, similar to the previous case, $a\rightarrow +\infty $ and $\rho _{\mathrm{r}}\rightarrow 0$ as $ t\rightarrow +\infty $, but, in the early universe, there is a finite maximum value that energy density can reach with $\rho =\rho _{\mathrm{r,max}}=-\frac{1}{8\alpha }$ when the length of the cosmic scale factor reaches its minimum as $a_{\mathrm{min}}=a_{1}\left( -\frac{12\alpha }{\kappa }\right) ^{\frac{1}{4}},$ and the Hubble parameter reaches its maximum as $H_{\mathrm{max}}=\frac{1}{4}\sqrt{-\frac{\kappa }{3\alpha }}$ at $t=t_{\mathrm{min}}=\sqrt{\frac{-12\alpha }{\kappa }}$. We note that $\dot{\rho}_{\mathrm{r}}=-\frac{3}{2\kappa }\left( 1+\frac{12\alpha }{\kappa t^{2}}\right) ^{-\frac{1}{2
}}\,t^{-3}$ is always negative in both cases, implying that the energy density decreases monotonically as $t$ increases, which in turn guarantees that the modified field equations Eq. \eqref{eq:rhopresprime} will be indistinguishable from the standard Friedmann equations of a radiation dominated universe for sufficiently large $t$ values. Hence, we would expect the deviation from the standard radiation dominated universe to be (in)significant (after)before a certain time in the history of the universe depending on the value of the parameter $\alpha $. However, we note that the time evolution of the Hubble parameter and the value of the deceleration parameter are the same as that of the standard radiation dominated universe in GR, namely, $H(t)=H_{\mathrm{std}}(t)=\frac{1}{2t}$ and $q=q_{\mathrm{std}}=1$, where $q=-1-\frac{\dot{H}}{H^{2}}$. On the other hand, we see from Eq. \eqref{eq:rhopresprime} that the value of the Hubble parameter $H$ for a given value of the energy density differs from the one in the standard radiation dominated universe, $H_{\mathrm{std}}=\sqrt{\frac{\kappa }{3}\rho_{\mathrm{r}}},$ as follows:
\begin{equation}
\label{eqn:Hcompare}
\frac{H(\rho_{\mathrm{r}})}{H_{\mathrm{std}}(\rho_{\mathrm{r}})}=\sqrt{1+4\alpha \rho_{\mathrm{r}}}\,\,.
\end{equation}
In addition, the value of the energy density $\rho _{\mathrm{r}}$ of the radiation for a given cosmic time $t$ differs from the one in the standard radiation dominated universe, $\rho _{\mathrm{r,std}}=\frac{3}{4\kappa t^{2}}$, as follows:
\begin{equation}
\frac{\rho_{\mathrm{r}}}{\rho_{\mathrm{r,std}}}=\frac{\kappa}{6\alpha}\left(\sqrt{1+\frac{12\alpha}{\kappa\,t^2}}-1\right)\,t^2.
\end{equation}

We note that the dynamics of the very early universe is significantly modified; for $\alpha >0$, we have $H(\rho _{\mathrm{r}})\sim 2\sqrt{\alpha\rho _{\mathrm{r}}}\,H_{\mathrm{std}}(\rho _{\mathrm{r}})$ when $4\alpha\rho_{\mathrm{r}}\gg 1$ and $\frac{\rho _{\mathrm{r}}}{\rho _{\mathrm{r,std}}}\rightarrow 0$ as $t\rightarrow 0$, and for $\alpha <0$, we have $H(\rho _{\mathrm{r}})\sim \frac{H_{\mathrm{std}}(\rho _{\mathrm{r}})}{\sqrt{2}}$ when $\rho _{\mathrm{r}}\sim \rho _{\mathrm{max}}$ and $\frac{\rho _{\mathrm{r}}}{\rho _{\mathrm{r,std}}}\rightarrow 2$ as $t\rightarrow t_{\mathrm{min}}$. On the other hand, as expected, we have $\rho _{\mathrm{r}}\sim \rho _{\mathrm{r,std}}$ and $H(\rho _{\mathrm{r}})\sim H_{\mathrm{std}}(\rho _{\mathrm{r}})$ for $|4\alpha\rho _{\mathrm{r}}|\ll 1$, i.e., for sufficiently large values of cosmic time $t$ for both cases.

We depict, considering the limits on $\alpha $ given in Eq. \eqref{constraint1} from NSs, in Fig. \ref{rho_h} Hubble parameter $H$ versus $\rho _{\mathrm{r}} $ and in Fig. \ref{rho_r}  $\rho_{\rm r}$ versus cosmic time $t$. We immediately see from both Figs. \ref{rho_h}-\ref{rho_r} that the modification in the dynamics of the early universe in the EMSG model with respect to the standard cosmology becomes apparent for the times $t\lesssim 10^{-4}\,\mathrm{s}$ and the energy density values $\rho _{\mathrm{r}}\gtrsim 10^{34}\,\mathrm{erg\,cm^{-3}}$ ($T\gtrsim 10^{12}\,\mathrm{K}$). In the positive $\alpha $ limit, i.e., $\alpha \sim 10^{-37}\,\mathrm{cm^{3}/erg}$, in contrast to the GR, we have $H\sim 4\kappa \alpha \rho _{\mathrm{r}}$ and $\rho _{\mathrm{r}}\sim \sqrt{\frac{3}{16\kappa \alpha }}\,\frac{1}{t}$. For the negative $\alpha $ limit, i.e., $\alpha \sim-10^{-38}\,\mathrm{cm^{3}/erg}$, the universe reaches the nonsingular minimum of the expansion scale factor, $a_{\min },$ before significant deviations in the values of $H(\rho _{\mathrm{r}})$ and $\rho _{\mathrm{r}}(t)$ develop with respect to the GR values. Thus, the values of these parameters in GR and EMSG are of the same order of magnitude at $t=t_{\mathrm{min}}$, although in GR this time is not an expansion minimum and the universe heads towards a big bang singularity as $t$ keeps on decreasing to $t=0$ from $t=t_{\mathrm{min}}$. Considering the lower limit of the constraints given in Eq. \eqref{constraint1}, i.e. $\alpha =-10^{-38}\,\mathrm{cm^{3}\,erg^{-1}}$, there is a change to the evolution of the very early universe qualitatively$-$namely, there is no initial singularity$-$because $\alpha $ can be negative. Hence, with this constraint we find that $\rho _{\mathrm{r,max}}\gtrsim 1.25\times 10^{37}\,\mathrm{erg\,cm^{-3}}$, the time of the beginning is $t_{\mathrm{min}}\lesssim 8.02\times 10^{-6}\,\mathrm{s}$, and the minimum size of the universe is $a_{\mathrm{min}}\lesssim 10^{-4}a_{1}$. To present a clearer comparison between the dynamics of the early universe in EMSG and GR, we also depict in Fig. \ref{devfromGRr} the evolution of $\frac{\rho_{\mathrm{r}}}{\rho _{\mathrm{r,std}}}$ in cosmic time $t$ for the limits on $\alpha $ given in Eq. \eqref{constraint1} from NSs.

\begin{table*}[t]
\centering{\footnotesize
\captionsetup{justification=raggedright,singlelinecheck=false}
\begin{tabular}{lllllll}
\hline\hline
&  &  &  &  &  &    \\
& Standard & Standard & Standard & $\alpha=10^{-37}$ & $\alpha=-10^{-38}$ & $ |\alpha|\lesssim10^{-48}$  \\
Event & Energy-scale & En.density-scale & Time-scale & $\mathrm{cm^3/erg}$ &
$\mathrm{cm^3/erg}$ & $\mathrm{cm^3/erg}$ \\
\hline\hline
&  &  &  &  &  &    \\
Matter-radiation equality & $10^4$ K & $10^3\,\mathrm{erg\,cm^{-3}}$ & $10^{4}$ yr & $10^{4}$ yr & $10^{4}$ yr &$10^{4}$ yr  \\[6pt]
Primordial nucleosynthesis & $10^9$ K & $10^{22}\,\mathrm{erg\,cm^{-3}}$ & $10^{2}$ s & $10^{2}$ s & $10^{2}$ s & $10^{2}$ s  \\[6pt]
$\nu$ decoupling, $e\bar{e}$ annihilation & $10^{10}$ K & $10^{26}\,\mathrm{erg\,cm^{-3}}$ & $1$ s & $1$ s & $1$ s & $1$ s  \\[6pt]
$\gamma$, $\nu$, $e$, $\bar{e}$, $n$ and $p$ thermal equilibrium & $10^{11}$ K & $10^{30}\,\mathrm{erg\,cm^{-3}}$ & $10^{-2}$ s & $10^{-2}$ s & $10^{-2}$ s & $10^{-2}$ s  \\[6pt]
Quark-hadron phase transition & $10^{12}$ K & $10^{34}\,\mathrm{erg\,cm^{-3}} $ & $10^{-4}$ s & $10^{-4}$ s & $10^{-4}$ s & $10^{-4}$ s  \\[6pt]
Electroweak phase transition & $10^{15}$ K & $10^{46}\,\mathrm{erg\,cm^{-3}}$ & $10^{-10}$ s & $10^{-16}$ s & N/A & $10^{-10}$ s  \\[6pt] \hline\hline
\end{tabular}
}
\caption{The time scales of some important energy-scales are calculated. We consider the corresponding energy and time scales of some key events in the standard cosmology and compare with our calculated values.}
\label{tab:keyevents}
\end{table*}

We presented, in \autoref{tab:keyevents}, the corresponding energy density and time scales of some key events in the standard cosmology and a rough comparison of the relevant time scales in standard cosmology and cosmology in EMSG\footnote{ For the times after the matter-radiation equality, radiation is negligible and then cosmological field equations read $3H^{2}=\kappa \rho _{\mathrm{m}}+\kappa \alpha \rho _{\mathrm{m}}^{2}+\Lambda $ and $-2\dot{H}-3H^{2}=\kappa \alpha \rho _{\mathrm{m}}^{2}-\Lambda $, where $\rho _{\mathrm{m}}=\rho _{\mathrm{m,0}}a^{-3}$ since the conservation equation holds for $w_{\mathrm{m}}=0$ in EMSG \cite{aka17,barrow17}. It is noteworthy that these are in the form that appear in braneworld cosmology \cite{BraneW1} for $\alpha >0$ and loop quantum cosmology \cite{LQC11} for $\alpha <0$ and that the corresponding energy density and pressure for the new terms arise from EMSG for matter source can effectively be written as $p_{\mathrm{s}}=\rho _{\mathrm{s}}=\alpha \rho _{\mathrm{m}}^{2}$ in GR, namely, they contribute to the Einstein's field equations like a stiff fluid \cite{zel61} that changes as $\rho _{\mathrm{s}}=\rho _{\mathrm{s,0}}a^{-6}$ due to its EoS parameter $w_{\mathrm{s}}=1$. However, here there is a specific relation $\rho _{\mathrm{s,0}}=\alpha \rho _{\mathrm{m,0}}^{2}$, which implies that $\rho _{\mathrm{s}}>0$ for $\alpha >0$ and $\rho _{\mathrm{s}}<0$ for $\alpha <0$. One may see Ref. \cite{Chavanis15} for a comprehensive investigation of the inclusion of a stiff fluid (for either $\rho _{\mathrm{s}}>0$ or $\rho _{\mathrm{s}}<0$) into $\Lambda $CDM model and the exact explicit solutions, which can be straightforwardly adapted to our model by keeping in mind that $\rho _{\mathrm{s,0}}=\alpha \rho _{\mathrm{m,0}}^{2}$. Accordingly, using $\rho _{\mathrm{s,0}}=\alpha \rho _{\mathrm{m,0}}^{2}$ with $\rho _{\mathrm{m,0}}\sim 10^{-9}\,\mathrm{erg\,cm^{-3}}$ for the present universe, and the constraints on $\alpha $ given in Eq. \eqref{constraint1}, we find that $-10^{-47}\lesssim \rho _{\mathrm{s,0}}/\rho _{\mathrm{m,0}}=\alpha \rho _{\mathrm{m,0}}\lesssim 10^{-46}$. Hence, our model for the times after matter-radiation equality would obviously be indistinguishable from $\Lambda  $CDM model and for this reason we do not elaborate on the late universe for the sake of brevity.} considering the limits on $\alpha $ given in Eq. \eqref{constraint1} from NSs. We see that time scales of the relevant energy density-scales do not differ from that of the standard cosmology up to the energy density-scales relevant to quark-hadron phase transition. At higher energy density-scales, on the other hand, we see a considerable deviation in the time scales, namely, for the positive boundary $\alpha =10^{-37}\,\mathrm{cm^{3}/erg}$ given in Eq. \eqref{constraint1}, energy density-scales relevant to the electroweak phase transition is reached when $t\sim10^{-16}\,\mathrm{s}$ while it is $t\sim 10^{-10}\,\mathrm{s}$ in standard cosmology, and for the negative boundary $\alpha =-10^{-38}\,\mathrm{cm^{3}/erg}$ given in Eq. \eqref{constraint1} these energy density-scales would never be reached. We also showed in the last column that if $|\alpha|\lesssim 10^{-48}\,\mathrm{cm^{3}/erg}$ then there would be no significant deviation from the standard cosmology up to the energy density scales relevant to the electroweak phase transition $10^{46}\,\mathrm{erg\,cm^{-3}}$.

We would like to end this section by a brief remark on the primordial nucleosynthesis in EMSG under the constraints given in Eq. \eqref{constraint1} obtained from NSs. The standard BBN and the well known phases of the universe that precede BBN (such as $\gamma$, $\nu$, $e$, $\bar{e}$, $n$ and $p$ thermal equilibrium; $\nu$ decoupling, $e\bar{e}$ annihilation; primordial nucleosynthesis) take place in the period of time from $\sim10^{-2}\,\mathrm{s}$ to $\sim10^2\,\mathrm{s}$ and during which the universe is radiation dominated and energy density scale drops from $\rho_{\mathrm{r}}\sim10^{22}\,\mathrm{erg\,cm^{-3}}$ to $\rho_{\mathrm{r}}\sim10^{29}\,\mathrm{erg\,cm^{-3}}$. Accordingly, during that period the modification term in the Hubble parameter, Eq. \eqref{eq:rhopresprime}, due to EMSG changes as from $-10^{-15}\lesssim 4\alpha\rho_{\mathrm{r}} \lesssim 10^{-14}$ to $-10^{-8}\lesssim 4\alpha\rho_{\mathrm{r}} \lesssim 10^{-7}$. These imply almost no deviation from the standard cosmology Hubble expansion rate at the time- and energy density-scales relevant to primordial nucleosynthesis, namely, we have $-10^{-15} \lesssim H^2(\rho_{\mathrm{r}})/H^2_{\mathrm{std}}(\rho_{\mathrm{r}})-1 \lesssim 10^{-14}$ and $-10^{-8} \lesssim H^2{(\rho_{\mathrm{r}})}/H^2_{\mathrm{std}}(\rho_{\mathrm{r}})-1 \lesssim 10^{-7}$, respectively. Hence, under the constraints given in Eq. \eqref{constraint1}, because energy density, time-energy density relation and Hubble expansion rate remain unaltered in EMSG, it is conceivable that BBN processes would remain the same as in the standard BBN. However, our conclusion here is subject to the thorough analysis of primordial nucleosynthesis in EMSG and will be presented elsewhere.

%%%%%%%%%%%%%%%%%%%%%%%%%%%%%%%%%%%%%%%%%%%%%%%%%%%%%%%%%%%%

\section{Concluding remarks}

\label{sec:discuss}
%%%%%%%%%%%%%%%%%%%%%%%%%%%%%%%%%%%%%%%%%%%%%%%%%%%%%%%%%%%%

In this study, we have tested energy-momentum squared gravity (EMSG) model in the strong gravity field regime using neutron stars (NSs). We have discussed further features of the EMSG model on theoretical and observational grounds, identifying the energy density scales at which EMSG differs significantly from standard GR. We also showed that the modifications to GR in EMSG are effective at relatively high energy densities and would lead to new effects in the early universe or in compact astrophysical objects. Therefore black holes (BH's) and NSs in EMSG can in principal have features that make them observationally distinguishable from their standard GR counterparts. As stated above, considering the parameter space for quantifying the strength of a gravitational field, the strongest gravitational fields around astrophysical systems can be found near NSs and BHs in x-ray binaries. In Ref. \cite{ros16}, the authors examined charged BH (Kerr-Newman) solutions within the framework of EMSG since it is obvious that if the matter energy density is zero, i.e., in the vacuum, EMSG is equivalent to GR hence the forms of Kerr, Schwarzschild and de Sitter solutions in the framework of EMSG will be exactly the same as in GR. Therefore, since observed astronomical objects do not possess an appreciable net electric charge, astrophysical BH's are neutral, NSs remain the prime site for testing the deviations of EMSG from GR in the strong field regime. New ways of using data on NS properties to test modified gravity theories have recently been proposed (see e.g. Ref. \cite{don17} for a recent review) that exploit correlations between different observables identified across NS populations.

We obtained the hydrostatic equilibrium equations in spherical symmetry from the field equations within the framework of EMSG. We discussed the local stability of the hydrostatic equilibrium of a mass distribution as well as the stability of stellar configurations with respect to small perturbations and place on $\alpha$, the free parameter that determines the coupling strength of the EMSG modification, some preliminary theoretical estimations and constraints, which then we employed for sorting out the stable solutions in our numerical calculations. We solved the hydrostatic equilibrium equations numerically for four realistic equations of state (EoSs) that describe the dense matter inside NSs and obtained the mass-radius relations for each of them depending on the value of $\alpha $. We have also constrained the value of $\alpha $ by comparing the computed mass-radius relations with the recent observational measurements of those for actual NSs. We have determined the maximum mass of NS for each EoS depending on the value of $\alpha $. We have discussed the nontrivial influence of EMSG on NS configurations due to its effective stiffening and softening of the EoS within the NS, depending on $\alpha $. We have shown the presence of a critical value of the EoS parameter, $P/\rho =1/3,$ around which an EoS is effectively stiffened and softened. Any EoS experiencing this critical value would lead to the presence of domains within a NS where the EoS is effectively stiffened and softened. We presented some insights into how the interplay of these domains would lead to the complicated behaviour of the maximum mass depending on the value of $\alpha $.

We have seen that the hyperonic EoS named GM1 Y4, which is ruled out within GR since it predicts $M_{\max }<2\,M_{\odot }$, can lead to maximum masses increasing with $\alpha $, finally exceeding $2M_{\odot }$ for $\alpha \simeq 0.68\times 10^{-37}\,\mathrm{cm^{3}\,erg^{-1}}$. Yet we concluded that this is not a satisfactory resolution of this so called \textquotedblleft hyperon puzzle\textquotedblright\ as for such values of $\alpha $ the model predicts very large radii $\sim 15\,\mathrm{km}$ -- somewhat greater than the observational bounds. On the other hand, we have seen that for some values of $\alpha $, compatibility of SkOp and MS2 with the mass-radius observations were improved compared to GR. We have concluded that the APR, SkOp, and MS2 EoS parametrizations, which are already compatible with observations within GR, would still be compatible with observations for the range $-10^{-38}\,\mathrm{cm^{3}/erg}\lesssim \alpha \lesssim 10^{-37}\,\mathrm{cm^{3}/erg}$ thus placing an order-of-magnitude constraint on the value of $\alpha$  \footnote{ We have allowed the free parameter $\alpha $ of EMSG to take both positive and negative values. However, in a work under progress it is found that the rate of change of the entropy per baryon with respect to time is negative for $\alpha <0$ in EMSG, which suggests a thermodynamic problem for that sign. Similarly, the effective EoS stiffer than Zeldovich fluid ($P/\rho=1$) that can be achieved when $\alpha $ is negative in the present paper might be signalling some stability issues for $\alpha<0 $. If it turns out in future works that $\alpha <0$ indeed leads to some unrealistic physical results, then the range of values allowed for $\alpha $ we obtain in this work would be further restricted as $0\leq \alpha \lesssim 10^{-37}\,\mathrm{cm^{3}/erg}$.}. The degeneracies between the EoS and gravity do not allow for a precise constraint, yet this is still much tighter than any solar system test could provide. Finally, in the cosmological context, we also showed that, under these constraints, there would be no significant deviation from the standard cosmology up to the energy density and time scales, $\sim 10^{34}\,\mathrm{erg\,cm^{-3}}$ and $\ t\simeq 10^{-4}\,\mathrm{s}$, led us to conclude that EMSG leaves the most important features of standard cosmology such as the standard big bang nucleosynthesis unaltered, but yet it may still have far reaching consequences for the dynamics of the very early universe relevant to issues like inflation, early domination by spatial anisotropy, cosmological bounce, and the initial singularity.

\acknowledgements We would like to thank Charles~V.R.~Board for discussions. \"{O}.A. acknowledges the support by the Science Academy in scheme of the Distinguished Young Scientist Award (BAGEP). N.K. acknowledges the post-doctoral research support she is receiving from the {\.{I}}stanbul Technical University. J.D.B. is supported by the Science and Technology Facilities Council (STFC) of the UK.\\
\newpage

\appendix

\section{Stability of relativistic fluid configurations in EMSG}
\label{sec:apndx0}

We present an analysis of the stability of relativistic stars in equilibrium in EMSG with respect to small perturbations. We follow the procedure given in \citep{cha64a,cha64b} for the stability of spherical fluid configurations in GR. Accordingly, we apply time dependent radial perturbations to the static solutions and examine the frequency of the perturbations.

In the case of the radial perturbations, the metric is written as
\begin{equation}
\begin{aligned}
\dif s^2=&-e^{2\nu(r)}\left(1+h\left(r,t\right)\right)\dif t^2+e^{2\lambda(r)}\left(1+f\left(r,t\right)\right)\dif r^2\\
&+r^2\dif\theta^2+r^2\sin^2\theta\dif\phi^2
\end{aligned}
\end{equation}
and the components of four velocity are given as
\begin{equation}
\begin{aligned}
u_0&=-e^{\nu(r)}\left(1+\frac12h(r,t)\right),\\
u_1&=e^{2\lambda(r)-\nu(r)}v(r,t) \quad\textnormal{and}\quad u_3=u_4=0,
\end{aligned}
\end{equation}
where $v$, $f$ and $h$ are perturbations. The density and the pressure are replaced by
\begin{align}
\varrho\left(r,t\right)=&\rho\left(r\right)+\delta\rho\left(r,t\right), \\
\mathcal{P}\left(r,t\right)=&P\left(r\right)+\delta P\left(r,t\right),
\end{align}
where $\rho$ and $P$ denote solutions of the hydrostatic equations given in Eqs. \eqref{TOV1}-\eqref{TOV2}, $\delta\rho$ and $\delta P$ denote perturbations of solutions of the hydrostatic equations. Accordingly, the field equations Eq. \eqref{fieldeq2} read
\begin{align}
\frac{e^{-2\lambda}}{r^2}
\left(-1+e^{2\lambda}+2r\frac{\dif\lambda}{\dif r} \right)&=\kappa \rho_{\rm eff}, \label{Eq30np}
\\
\frac{1}{r^2}\frac{\partial}{\partial r}\left(rfe^{-2\lambda}\right)
&= \kappa\delta\rho_{\rm eff}, \label{Eq30}
\\
\frac{e^{-2\lambda}}{r^2}\left(1-e^{2\lambda}+2r\frac{\dif\nu}{\dif r}\right)&=\kappa P_{\rm eff}, \label{Eq31np}
\\
\frac{e^{-2\lambda}}{r}\frac{\partial h}{\partial r}-f\frac{e^{-2\lambda}}{r^2}\left(1+2r\frac{\dif\nu}{\dif r}\right)&=\kappa \delta P_{\rm eff}, \label{Eq31}
\end{align}
where only linear terms of perturbations are kept and the effective density and pressure are defined as
\begin{widetext}
\begin{align}
\rho_{\rm eff}\left(r\right)&=\rho\left(r\right)+\alpha\left[\rho\left(r\right)^2+8\rho\left(r\right)P\left(r\right)+3P\left(r\right)^2\right],
\\
\delta\rho_{\rm eff} \left(r,t\right)&=\delta\rho(r,t)+\alpha\left[2\rho(r)\delta\rho(r,t)+8\delta\rho(r,t)P(r)+8\rho(r)\delta P(r,t)+6P(r)\delta P(r,t)\right],
\\
P_{\rm eff} \left(r\right)&= P\left(r\right)+\alpha\rho\left(r\right)^2+3\alpha P\left(r\right)^2,
\\
\delta P_{\rm eff} \left(r,t\right)&= \delta P\left(r,t\right)+2\alpha\rho\left(r\right)\delta \rho\left(r,t\right)+6\alpha P\left(r\right)\delta P\left(r,t\right).
\end{align}
\end{widetext}

The ``$t$-$r$'' component of the field equations reads
\begin{equation}
\frac{\partial f}{\partial t}=-\kappa re^{2\lambda}
\left(\rho_{\rm eff}+P_{\rm eff}\right)v, \label{Eq32}
\end{equation}
and $\nabla^\mu G_{\mu 1}=0$ implies
\begin{align}
0=&
\left( \rho_{\rm eff}+P_{\rm eff}\right)\frac{\dif\nu}{\dif r}
+\frac{\dif P_{\rm eff}}{\dif r}, \label{Eq33e}
\\
0=&\left(\delta \rho_{\rm eff}+\delta P_{\rm eff}\right)\frac{\dif\nu}{\dif r}+\frac{1}{2}\left(\rho_{\rm eff}+P_{\rm eff}\right)\frac{\partial h}{\partial r}+\frac{\partial \delta P_{\rm eff}}{\partial r}  \label{Eq33}\\
&+e^{2\lambda-2\nu}\left(\rho_{\rm eff}+P_{\rm eff}\right)\frac{\partial v}{\partial t}.\nonumber
\end{align}
Using the Lagrangian displacement $\xi$ introduced as
\begin{equation}
v=\frac{\partial\xi}{\partial t},
\label{Ld}
\end{equation}
in the integration of Eq. \eqref{Eq32} we obtain
\begin{equation}
f=-\kappa re^{2\lambda}
\left(\rho_{\rm eff}+P_{\rm eff}\right)\xi.\label{Eq36}
\end{equation}
Using this in Eq. \eqref{Eq30} and Eq. \eqref{Eq31} we reach
\begin{align}
\delta \rho_{\rm eff}=&
-\frac{1}{r^2}\frac{\partial}{\partial r}\left[r^2\left(\rho_{\rm eff}+P_{\rm eff}\right)\xi \right],
\\
\frac{e^{-2\lambda}}{r}\frac{\partial h}{\partial r}
=&-\kappa
\left(\rho_{\rm eff}+P_{\rm eff}\right)\left(\frac{1}{r}+2\frac{\dif\nu}{\dif r}\right)\xi+
\kappa \delta P_{\rm eff},
\label{Eq40}
\end{align}
respectively. Addition of the leading terms of the ``$t$-$t$'' [Eq. \eqref{Eq30np}] and ``$r$-$r$'' [Eq. \eqref{Eq31np}] components of the field equations side by side gives
\begin{equation}
2\frac{e^{-2\lambda}}{r}\frac{\dif}{\dif r}\left(\lambda+\nu\right)=\kappa\left(\rho_{\rm eff}+P_{\rm eff}\right).
\end{equation}
Using this, Eq. \eqref{Eq40} can be written as
\begin{equation}
\begin{aligned}
&\frac{1}{2}\left(\rho_{\rm eff}+P_{\rm eff}\right)\frac{\partial h}{\partial r}=\\
&\quad\left[\delta P_{\rm eff}-\left(\rho_{\rm eff}+P_{\rm eff}\right)\left(\frac{1}{r}+2\frac{\dif\nu}{\dif r}\right)\xi
\right]\frac{\dif}{\dif r}\left(\lambda+\nu\right). \label{Eq41}
\end{aligned}
\end{equation}
Introducing the time dependence of all perturbations as $\exp\left(i\sigma t\right)$ and using the relations obtained above, Eq. \eqref{Eq33} can now be written as
\begin{equation}
\begin{aligned}
0=&
-\frac{1}{r^2}\frac{\dif}{\dif r}\left[r^2\left(\rho_{\rm eff}+P_{\rm eff}\right)\xi \right]\frac{\dif\nu}{\dif r}
+\delta P_{\rm eff}\frac{\dif}{\dif r}\left(\lambda+2\nu\right)\\
&-\left(\rho_{\rm eff}+P_{\rm eff}\right)\left(\frac{1}{r}+2\frac{\dif\nu}{\dif r}\right)\xi
\frac{\dif}{\dif r}\left(\lambda+\nu\right)
+\frac{\dif \delta P_{\rm eff}}{\dif r}\\
&-\sigma^2e^{2\lambda-2\nu}\left(\rho_{\rm eff}+P_{\rm eff}\right)\xi.
\label{Eq43}
\end{aligned}
\end{equation}
Next, because baryon number is $\mathcal{N}\equiv \mathcal{N}\left(\varrho_{\rm eff},\mathcal{P}_{\rm eff}\right)$, the conservation of baryon number, $\nabla_k\left(\mathcal{N}u^{k}\right)=0$, using $\mathcal{N}=N(r)+\delta N(r,t)$, leads to
\begin{align}
\delta P_{\rm eff}=
-\xi\frac{\dif P_{\rm eff}}{\dif r}
-\gamma_{\rm eff} P_{\rm eff}\frac{e^\nu}{r^2}\frac{\partial}{\partial r}\left(r^2\xi e^{-\nu}\right),
\label{Eq53}
\end{align}
where
\begin{equation}
\gamma_{\rm eff}=\frac{1}{\mathcal{P}_{\rm eff}\partial \mathcal{N}/\partial \mathcal{P}_{\rm eff}}\left[\mathcal{N}-\left(\varrho_{\rm eff}+\mathcal{P}_{\rm eff}\right)\frac{\partial \mathcal{N}}{\partial \varrho_{\rm eff}}\right]
\end{equation}
is the effective ratio of the specific heats.

Finally, using Eq. \eqref{Eq53} in Eq. \eqref{Eq43} and employing Eqs. \eqref{Eq33e}, \eqref{Eq30np} and \eqref{Eq31np}, we reach the following Sturm-Liouville equation \citep{dat+92}, the eigenvalue equation governing radial oscillations of a spherical star in our model,
\begin{equation}
-\sigma^2 \omega\left(r\right)r^2e^{-\nu}\xi=q\left(r\right)r^2e^{-\nu}\xi
+\frac{\dif}{\dif r}\left[
k\left(r\right)\frac{\dif}{\dif r}\left(r^2e^{-\nu}\xi\right)\right],
\end{equation}
where
\begin{align}
\omega\left(r\right)=\frac{1}{r^2}e^{3\lambda+\nu}\left(\rho_{\rm eff}+P_{\rm eff}\right),\quad k\left(r\right)= e^{\lambda+3\nu}\frac{\gamma_{\rm eff} P_{\rm eff}}{r^2} \label{eqn:fundamode}
\end{align}
and
\begin{equation}
\begin{aligned}
q\left(r\right)=-\frac{e^{\lambda+3\nu}}{r^2}
\bigg[\frac{4}{r}\frac{\dif P_{\rm eff}}{\dif r}+&\kappa e^{2\lambda}P_{\rm eff}\left(P_{\rm eff}+\rho_{\rm eff}\right)\\
&-\frac{1}{P_{\rm eff}+\rho_{\rm eff}}\left(\frac{\dif P_{\rm eff}}{\dif r}\right)^2
\bigg].
\end{aligned}
\end{equation}
We note that the Sturm-Liouville equation in EMSG is exactly the same with the one in GR, except that the quantities $\rho$ and $P$ in GR are replaced by the effective quantities $\rho_{\rm eff}=\rho_{\rm eff}(\alpha)$ and $P_{\rm eff}=P_{\rm eff}(\alpha)$ with  $\rho=\rho_{\rm eff}(\alpha=0)$ and $P=P_{\rm eff}(\alpha=0)$.  Provided that $k>0$ and $\omega>0$, the eigenvalues of the Sturm-Liouville equation are all real and eigenvalues form an infinite discrete sequence $\sigma_0^2<\sigma_1^2<\sigma_2^2<\dots$ (subscripts denote node numbers).  The conditions $k>0$ and $\omega>0$ are guaranteed to be satisfied in GR (the case $\alpha=0$ in EMSG) since $\rho>0$ and $P>0$ within stars by definition. In EMSG, on the other hand, we see from Eq. \eqref{eqn:fundamode} that $k>0$ and $\omega>0$ require $\gamma_{\rm eff} P_{\rm eff}>0$ and $\rho_{\rm eff}+P_{\rm eff}>0$, respectively, and hence are subject to the free parameter $\alpha$ (for instance, sufficiently large negative values of $\alpha$ may lead to negative $P_{\rm eff}$ values). We note that the condition ${\rm d}m/{\rm d}r>0$ we employ in this work/numerical solutions already ensures $\rho_{\rm eff}>0$ from Eq. \eqref{TOV1}. We demand $P_{\rm eff}>0$, which in turn together with $\rho_{\rm eff}>0$ constrain $\gamma_{\rm eff}$ as $P>0$ and $\rho>0$ constrain $\gamma$ in GR.  Thus, demanding $P_{\rm eff}>0$ in our work/numerical solutions in addition to the condition ${\rm d}m/{\rm d}r>0$ ensuring $\rho_{\rm eff}>0$ we can look for stable neutron star solutions by considering static stability criterion given in Eq. \eqref{stability_star} (which is necessary but not sufficient ) as well as the sufficient criterion which enables one to determine the precise number of unstable normal radial modes using the $M(R)$ curve as it is done in GR case (\citep[see Sec. 6.5 of Ref.][for further discussion]{han+07}. Accordingly, as further discussed in Sec. \ref{sec:stability}, we demand our numerical solutions to satisfy $\rho_{\rm eff} =\rho+\rho_{\rm EMSG}>0$ and $P_{\rm eff}= P+P_{\rm EMSG}>0$, thereby we are able to decide whether the solutions of NSs in equilibrium presented in this paper are stable or not with respect to \textit{any} small perturbations. Note that the stability of a nonrotating star, as it is the case in our work, with respect to small \textit{radial} perturbations implies the stability with respect to \textit{any} small perturbations of the star (see Ref. \cite{han+07} for details).

\section{Stability of the solution for the radiation-dominated universe}
\label{sec:apndx}

The cosmological field equations for the radiation-dominated universe given in Eq. \eqref{eq:rhopresprime} satisfy the continuity equation
\begin{equation}
\dot{\rho}_{\mathrm{r}}+4H\frac{1+4\alpha\rho_{\mathrm{r}}}{1+8\alpha\rho_{\mathrm{r}}}\rho_{\mathrm{r}}=0,
\end{equation}
and the following background solution given in Eq. \eqref{bcksol}
\begin{equation}
H(t)={\frac{1}{2t}}\,\,\textnormal{and}\,\,\rho _{\mathrm{r}}(t)=\frac{1}{8\alpha }\left( \sqrt{1+\frac{12\alpha }{\kappa t^{2}}}-1\right) ,
\label{radsols}
\end{equation}
where $a_{1}$ is the length of the cosmic scale factor when the age of the universe is $t=1\,\mathrm{s}$. We check the stability of this solution considering linear perturbations $\delta (t)$ and $\delta _{r}(t)$ about the backgrounds $H(t)$ and $\rho _{\mathrm{r}}(t)$ as
\begin{equation}
\label{perturb}
\mathcal{H}(t)=H(t)\left[1+\delta(t)\right] \,\, \textnormal{and}\,\, \varrho_{\mathrm{r}}(t)=\rho_{\mathrm{r}}(t)\left[1+\delta_r(t)\right],
\end{equation}
respectively. The perturbed modified Friedmann and continuity equations are then given by
\begin{align}
6H(t)\delta (t)=\kappa \,\delta _{\mathrm{r}}(t)\,\rho _{\mathrm{r}}(t)\left[1+8\alpha \rho _{\mathrm{r}}(t)\right] ,  \label{pert1} \\
\dot{\delta _{\mathrm{r}}}(t)+4H(t)\frac{1+4\alpha \rho _{\mathrm{r}}(t)}{1+8\alpha \rho _{\mathrm{r}}(t)}\,\delta (t) =0.
\label{pert2}
\end{align}
Substituting $\delta (t)$ from Eq. \eqref{pert1} into Eq. \eqref{pert2}, and then solving the resultant equation we find
\begin{equation}
\begin{aligned}
\label{pertsol}
\delta_{\rm r}(t)=c_1 \,{\rm exp}\int-\frac{2}{3}\frac{\kappa \rho_{\rm r}(t)\left[1+4\alpha\rho_{\rm
r}(t)\right]}{H(t)} {\rm d}t.
\end{aligned}
\end{equation}
Next, using this in Eq. \eqref{pert2}, we find
\begin{align}
\delta(t)=&\frac{\kappa \rho_{\mathrm{r}}(t)\left[1+8\alpha\rho_{\mathrm{r}}(t)\right]}{6H(t)^2}  \notag \\
& \times c_1\,\mathrm{exp}\int -\frac{2}{3}\frac{\kappa \rho_{\mathrm{r}}(t)\left[1+4\alpha\rho_{\mathrm{r}}(t)\right]}{H(t)} \mathrm{d}t.
\label{pertsol2}
\end{align}

Finally, using the background solutions Eq. \eqref{radsols} (i.e., Eq. \eqref{bcksol} in the main text) with these, we find the following solution for the linear perturbations:
\begin{equation}
\delta (t)=\frac{c_{1}}{t}\frac{\sqrt{1+\frac{12\alpha }{\kappa t^{2}}}}{1+\sqrt{1+\frac{12\alpha }{\kappa t^{2}}}}\quad \textnormal{and}\quad \delta _{\mathrm{r}}(t)=\frac{c_{1}}{t}.
\label{pertsol3}
\end{equation}
We see that our solution is stable against linear perturbations since both $\delta (t)$ and $\delta _{\mathrm{r}}(t)$ decrease to zero monotonically as the cosmic time $t$ grows for all cases that we are interested in, i.e., for $\alpha >0$, $\alpha =0$ (corresponding to GR) and $\alpha <0$ (with $t\geq t_{\mathrm{min}}=\sqrt{-12\alpha /\kappa }$).

\bibliography{refs}

\begin{thebibliography}{72}%
\makeatletter
\providecommand \@ifxundefined [1]{%
 \@ifx{#1\undefined}
}%
\providecommand \@ifnum [1]{%
 \ifnum #1\expandafter \@firstoftwo
 \else \expandafter \@secondoftwo
 \fi
}%
\providecommand \@ifx [1]{%
 \ifx #1\expandafter \@firstoftwo
 \else \expandafter \@secondoftwo
 \fi
}%
\providecommand \natexlab [1]{#1}%
\providecommand \enquote  [1]{``#1''}%
\providecommand \bibnamefont  [1]{#1}%
\providecommand \bibfnamefont [1]{#1}%
\providecommand \citenamefont [1]{#1}%
\providecommand \href@noop [0]{\@secondoftwo}%
\providecommand \href [0]{\begingroup \@sanitize@url \@href}%
\providecommand \@href[1]{\@@startlink{#1}\@@href}%
\providecommand \@@href[1]{\endgroup#1\@@endlink}%
\providecommand \@sanitize@url [0]{\catcode `\\12\catcode `\$12\catcode
  `\&12\catcode `\#12\catcode `\^12\catcode `\_12\catcode `\%12\relax}%
\providecommand \@@startlink[1]{}%
\providecommand \@@endlink[0]{}%
\providecommand \url  [0]{\begingroup\@sanitize@url \@url }%
\providecommand \@url [1]{\endgroup\@href {#1}{\urlprefix }}%
\providecommand \urlprefix  [0]{URL }%
\providecommand \Eprint [0]{\href }%
\providecommand \doibase [0]{http://dx.doi.org/}%
\providecommand \selectlanguage [0]{\@gobble}%
\providecommand \bibinfo  [0]{\@secondoftwo}%
\providecommand \bibfield  [0]{\@secondoftwo}%
\providecommand \translation [1]{[#1]}%
\providecommand \BibitemOpen [0]{}%
\providecommand \bibitemStop [0]{}%
\providecommand \bibitemNoStop [0]{.\EOS\space}%
\providecommand \EOS [0]{\spacefactor3000\relax}%
\providecommand \BibitemShut  [1]{\csname bibitem#1\endcsname}%
\let\auto@bib@innerbib\@empty
%</preamble>
\bibitem [{\citenamefont {Will}(2014)}]{wil14}%
  \BibitemOpen
  \bibfield  {author} {\bibinfo {author} {\bibfnamefont {C.~M.}\ \bibnamefont
  {Will}},\ }\bibfield  {title} {\enquote {\bibinfo {title} {The confrontation
  between general relativity and experiment},}\ }\href {\doibase
  10.12942/lrr-2014-4} {\bibfield  {journal} {\bibinfo  {journal} {Living
  Reviews in Relativity}\ }\textbf {\bibinfo {volume} {17}},\ \bibinfo {pages}
  {4} (\bibinfo {year} {2014})}\BibitemShut {NoStop}%
\bibitem [{\citenamefont {{Caldwell}}\ and\ \citenamefont
  {{Kamionkowski}}(2009)}]{cal09}%
  \BibitemOpen
  \bibfield  {author} {\bibinfo {author} {\bibfnamefont {R.~R.}\ \bibnamefont
  {{Caldwell}}}\ and\ \bibinfo {author} {\bibfnamefont {M.}~\bibnamefont
  {{Kamionkowski}}},\ }\bibfield  {title} {\enquote {\bibinfo {title} {{The
  Physics of Cosmic Acceleration}},}\ }\href {\doibase
  10.1146/annurev-nucl-010709-151330} {\bibfield  {journal} {\bibinfo
  {journal} {Annual Review of Nuclear and Particle Science}\ }\textbf {\bibinfo
  {volume} {59}},\ \bibinfo {pages} {397--429} (\bibinfo {year} {2009})},\
  \Eprint {http://arxiv.org/abs/0903.0866} {arXiv:0903.0866 [astro-ph.CO]}
  \BibitemShut {NoStop}%
\bibitem [{\citenamefont {{Sotiriou}}\ and\ \citenamefont
  {{Faraoni}}(2010)}]{sot10}%
  \BibitemOpen
  \bibfield  {author} {\bibinfo {author} {\bibfnamefont {T.~P.}\ \bibnamefont
  {{Sotiriou}}}\ and\ \bibinfo {author} {\bibfnamefont {V.}~\bibnamefont
  {{Faraoni}}},\ }\bibfield  {title} {\enquote {\bibinfo {title} {{f(R)
  theories of gravity}},}\ }\href {\doibase 10.1103/RevModPhys.82.451}
  {\bibfield  {journal} {\bibinfo  {journal} {Reviews of Modern Physics}\
  }\textbf {\bibinfo {volume} {82}},\ \bibinfo {pages} {451--497} (\bibinfo
  {year} {2010})},\ \Eprint {http://arxiv.org/abs/0805.1726} {arXiv:0805.1726
  [gr-qc]} \BibitemShut {NoStop}%
\bibitem [{\citenamefont {{de Felice}}\ and\ \citenamefont
  {{Tsujikawa}}(2010)}]{def10}%
  \BibitemOpen
  \bibfield  {author} {\bibinfo {author} {\bibfnamefont {A.}~\bibnamefont {{de
  Felice}}}\ and\ \bibinfo {author} {\bibfnamefont {S.}~\bibnamefont
  {{Tsujikawa}}},\ }\bibfield  {title} {\enquote {\bibinfo {title} {{f(R)
  Theories}},}\ }\href {\doibase 10.12942/lrr-2010-3} {\bibfield  {journal}
  {\bibinfo  {journal} {Living Reviews in Relativity}\ }\textbf {\bibinfo
  {volume} {13}},\ \bibinfo {pages} {3} (\bibinfo {year} {2010})},\ \Eprint
  {http://arxiv.org/abs/1002.4928} {arXiv:1002.4928 [gr-qc]} \BibitemShut
  {NoStop}%
\bibitem [{\citenamefont {{Nojiri}}\ and\ \citenamefont
  {{Odintsov}}(2011)}]{noj11}%
  \BibitemOpen
  \bibfield  {author} {\bibinfo {author} {\bibfnamefont {S.}~\bibnamefont
  {{Nojiri}}}\ and\ \bibinfo {author} {\bibfnamefont {S.~D.}\ \bibnamefont
  {{Odintsov}}},\ }\bibfield  {title} {\enquote {\bibinfo {title} {{Unified
  cosmic history in modified gravity: From F(R) theory to Lorentz non-invariant
  models}},}\ }\href {\doibase 10.1016/j.physrep.2011.04.001} {\bibfield
  {journal} {\bibinfo  {journal} {\physrep}\ }\textbf {\bibinfo {volume}
  {505}},\ \bibinfo {pages} {59--144} (\bibinfo {year} {2011})},\ \Eprint
  {http://arxiv.org/abs/1011.0544} {arXiv:1011.0544 [gr-qc]} \BibitemShut
  {NoStop}%
\bibitem [{\citenamefont {{Capozziello}}\ and\ \citenamefont {{de
  Laurentis}}(2011)}]{cap11}%
  \BibitemOpen
  \bibfield  {author} {\bibinfo {author} {\bibfnamefont {S.}~\bibnamefont
  {{Capozziello}}}\ and\ \bibinfo {author} {\bibfnamefont {M.}~\bibnamefont
  {{de Laurentis}}},\ }\bibfield  {title} {\enquote {\bibinfo {title}
  {{Extended Theories of Gravity}},}\ }\href {\doibase
  10.1016/j.physrep.2011.09.003} {\bibfield  {journal} {\bibinfo  {journal}
  {\physrep}\ }\textbf {\bibinfo {volume} {509}},\ \bibinfo {pages} {167--321}
  (\bibinfo {year} {2011})},\ \Eprint {http://arxiv.org/abs/1108.6266}
  {arXiv:1108.6266 [gr-qc]} \BibitemShut {NoStop}%
\bibitem [{\citenamefont {{Carroll}}\ \emph {et~al.}(2004)\citenamefont
  {{Carroll}}, \citenamefont {{Duvvuri}}, \citenamefont {{Trodden}},\ and\
  \citenamefont {{Turner}}}]{car+04}%
  \BibitemOpen
  \bibfield  {author} {\bibinfo {author} {\bibfnamefont {S.~M.}\ \bibnamefont
  {{Carroll}}}, \bibinfo {author} {\bibfnamefont {V.}~\bibnamefont
  {{Duvvuri}}}, \bibinfo {author} {\bibfnamefont {M.}~\bibnamefont
  {{Trodden}}}, \ and\ \bibinfo {author} {\bibfnamefont {M.~S.}\ \bibnamefont
  {{Turner}}},\ }\bibfield  {title} {\enquote {\bibinfo {title} {{Is cosmic
  speed-up due to new gravitational physics?}}}\ }\href {\doibase
  10.1103/PhysRevD.70.043528} {\bibfield  {journal} {\bibinfo  {journal}
  {\prd}\ }\textbf {\bibinfo {volume} {70}},\ \bibinfo {eid} {043528} (\bibinfo
  {year} {2004})},\ \Eprint {http://arxiv.org/abs/astro-ph/0306438}
  {astro-ph/0306438} \BibitemShut {NoStop}%
\bibitem [{\citenamefont {{Cognola}}\ \emph {et~al.}(2008)\citenamefont
  {{Cognola}}, \citenamefont {{Elizalde}}, \citenamefont {{Nojiri}},
  \citenamefont {{Odintsov}}, \citenamefont {{Sebastiani}},\ and\ \citenamefont
  {{Zerbini}}}]{cog+08}%
  \BibitemOpen
  \bibfield  {author} {\bibinfo {author} {\bibfnamefont {G.}~\bibnamefont
  {{Cognola}}}, \bibinfo {author} {\bibfnamefont {E.}~\bibnamefont
  {{Elizalde}}}, \bibinfo {author} {\bibfnamefont {S.}~\bibnamefont
  {{Nojiri}}}, \bibinfo {author} {\bibfnamefont {S.~D.}\ \bibnamefont
  {{Odintsov}}}, \bibinfo {author} {\bibfnamefont {L.}~\bibnamefont
  {{Sebastiani}}}, \ and\ \bibinfo {author} {\bibfnamefont {S.}~\bibnamefont
  {{Zerbini}}},\ }\bibfield  {title} {\enquote {\bibinfo {title} {{Class of
  viable modified f(R) gravities describing inflation and the onset of
  accelerated expansion}},}\ }\href {\doibase 10.1103/PhysRevD.77.046009}
  {\bibfield  {journal} {\bibinfo  {journal} {\prd}\ }\textbf {\bibinfo
  {volume} {77}},\ \bibinfo {eid} {046009} (\bibinfo {year} {2008})},\ \Eprint
  {http://arxiv.org/abs/0712.4017} {arXiv:0712.4017 [hep-th]} \BibitemShut
  {NoStop}%
\bibitem [{\citenamefont {{Psaltis}}(2008)}]{psa08}%
  \BibitemOpen
  \bibfield  {author} {\bibinfo {author} {\bibfnamefont {D.}~\bibnamefont
  {{Psaltis}}},\ }\bibfield  {title} {\enquote {\bibinfo {title} {{Probes and
  Tests of Strong-Field Gravity with Observations in the Electromagnetic
  Spectrum}},}\ }\href {\doibase 10.12942/lrr-2008-9} {\bibfield  {journal}
  {\bibinfo  {journal} {Living Reviews in Relativity}\ }\textbf {\bibinfo
  {volume} {11}},\ \bibinfo {eid} {9} (\bibinfo {year} {2008})},\ \Eprint
  {http://arxiv.org/abs/0806.1531} {arXiv:0806.1531} \BibitemShut {NoStop}%
\bibitem [{\citenamefont {{Psaltis}}(2009)}]{psa09}%
  \BibitemOpen
  \bibfield  {author} {\bibinfo {author} {\bibfnamefont {D.}~\bibnamefont
  {{Psaltis}}},\ }\bibfield  {title} {\enquote {\bibinfo {title} {{Two
  approaches to testing general relativity in the strong-field regime}},}\
  }\href {\doibase 10.1088/1742-6596/189/1/012033} {\bibfield  {journal}
  {\bibinfo  {journal} {Journal of Physics Conference Series}\ }\textbf
  {\bibinfo {volume} {189}},\ \bibinfo {eid} {012033} (\bibinfo {year}
  {2009})},\ \Eprint {http://arxiv.org/abs/0907.2746} {arXiv:0907.2746
  [astro-ph.HE]} \BibitemShut {NoStop}%
\bibitem [{\citenamefont {{Berti}}\ \emph {et~al.}(2015)\citenamefont
  {{Berti}}, \citenamefont {{Barausse}}, \citenamefont {{Cardoso}},
  \citenamefont {{Gualtieri}}, \citenamefont {{Pani}}, \citenamefont
  {{Sperhake}}, \citenamefont {{Stein}}, \citenamefont {{Wex}}, \citenamefont
  {{Yagi}}, \citenamefont {{Baker}}, \citenamefont {{Burgess}}, \citenamefont
  {{Coelho}}, \citenamefont {{Doneva}}, \citenamefont {{De Felice}},\ and\
  \citenamefont {{Ferreira}}}]{ber+15}%
  \BibitemOpen
  \bibfield  {author} {\bibinfo {author} {\bibfnamefont {E.}~\bibnamefont
  {{Berti}}}, \bibinfo {author} {\bibfnamefont {E.}~\bibnamefont {{Barausse}}},
  \bibinfo {author} {\bibfnamefont {V.}~\bibnamefont {{Cardoso}}}, \bibinfo
  {author} {\bibfnamefont {L.}~\bibnamefont {{Gualtieri}}}, \bibinfo {author}
  {\bibfnamefont {P.}~\bibnamefont {{Pani}}}, \bibinfo {author} {\bibfnamefont
  {U.}~\bibnamefont {{Sperhake}}}, \bibinfo {author} {\bibfnamefont {L.~C.}\
  \bibnamefont {{Stein}}}, \bibinfo {author} {\bibfnamefont {N.}~\bibnamefont
  {{Wex}}}, \bibinfo {author} {\bibfnamefont {K.}~\bibnamefont {{Yagi}}},
  \bibinfo {author} {\bibfnamefont {T.}~\bibnamefont {{Baker}}}, \bibinfo
  {author} {\bibfnamefont {C.~P.}\ \bibnamefont {{Burgess}}}, \bibinfo {author}
  {\bibfnamefont {F.~S.}\ \bibnamefont {{Coelho}}}, \bibinfo {author}
  {\bibfnamefont {D.}~\bibnamefont {{Doneva}}}, \bibinfo {author}
  {\bibfnamefont {A.}~\bibnamefont {{De Felice}}}, \ and\ \bibinfo {author}
  {\bibfnamefont {P.~G.}\ \bibnamefont {{Ferreira}}},\ }\bibfield  {title}
  {\enquote {\bibinfo {title} {{Testing general relativity with present and
  future astrophysical observations}},}\ }\href {\doibase
  10.1088/0264-9381/32/24/243001} {\bibfield  {journal} {\bibinfo  {journal}
  {Classical and Quantum Gravity}\ }\textbf {\bibinfo {volume} {32}},\ \bibinfo
  {eid} {243001} (\bibinfo {year} {2015})},\ \Eprint
  {http://arxiv.org/abs/1501.07274} {arXiv:1501.07274 [gr-qc]} \BibitemShut
  {NoStop}%
\bibitem [{\citenamefont {{Kat{\i}rc{\i}}}\ and\ \citenamefont
  {{Kavuk}}(2014)}]{kat14}%
  \BibitemOpen
  \bibfield  {author} {\bibinfo {author} {\bibfnamefont {N.}~\bibnamefont
  {{Kat{\i}rc{\i}}}}\ and\ \bibinfo {author} {\bibfnamefont {M.}~\bibnamefont
  {{Kavuk}}},\ }\bibfield  {title} {\enquote {\bibinfo {title}
  {{$f(R,T^{\mu\nu}T_{\mu\nu})$ gravity and Cardassian-like expansion as one of
  its consequences}},}\ }\href {\doibase 10.1140/epjp/i2014-14163-6} {\bibfield
   {journal} {\bibinfo  {journal} {European Physical Journal Plus}\ }\textbf
  {\bibinfo {volume} {129}},\ \bibinfo {eid} {163} (\bibinfo {year} {2014})},\
  \Eprint {http://arxiv.org/abs/1302.4300} {arXiv:1302.4300 [gr-qc]}
  \BibitemShut {NoStop}%
\bibitem [{\citenamefont {{Roshan}}\ and\ \citenamefont
  {{Shojai}}(2016)}]{ros16}%
  \BibitemOpen
  \bibfield  {author} {\bibinfo {author} {\bibfnamefont {M.}~\bibnamefont
  {{Roshan}}}\ and\ \bibinfo {author} {\bibfnamefont {F.}~\bibnamefont
  {{Shojai}}},\ }\bibfield  {title} {\enquote {\bibinfo {title}
  {{Energy-momentum squared gravity}},}\ }\href {\doibase
  10.1103/PhysRevD.94.044002} {\bibfield  {journal} {\bibinfo  {journal}
  {\prd}\ }\textbf {\bibinfo {volume} {94}},\ \bibinfo {eid} {044002} (\bibinfo
  {year} {2016})},\ \Eprint {http://arxiv.org/abs/1607.06049} {arXiv:1607.06049
  [gr-qc]} \BibitemShut {NoStop}%
\bibitem [{\citenamefont {Akarsu}\ \emph {et~al.}(2018)\citenamefont {Akarsu},
  \citenamefont {{Kat{\i}rc{\i}}},\ and\ \citenamefont {Kumar}}]{aka17}%
  \BibitemOpen
  \bibfield  {author} {\bibinfo {author} {\bibfnamefont {O.}~\bibnamefont
  {Akarsu}}, \bibinfo {author} {\bibfnamefont {N.}~\bibnamefont
  {{Kat{\i}rc{\i}}}}, \ and\ \bibinfo {author} {\bibfnamefont {S.}~\bibnamefont
  {Kumar}},\ }\bibfield  {title} {\enquote {\bibinfo {title} {{Cosmic
  acceleration in a dust only universe via energy-momentum powered gravity}},}\
  }\href {\doibase 10.1103/PhysRevD.97.024011} {\bibfield  {journal} {\bibinfo
  {journal} {Phys. Rev.}\ }\textbf {\bibinfo {volume} {D 97}},\ \bibinfo
  {pages} {024011} (\bibinfo {year} {2018})},\ \Eprint
  {http://arxiv.org/abs/1709.02367} {arXiv:1709.02367 [gr-qc]} \BibitemShut
  {NoStop}%
%%CITATION = ARXIV:1709.02367;%%
\bibitem [{\citenamefont {Board}\ and\ \citenamefont
  {Barrow}(2017)}]{barrow17}%
  \BibitemOpen
  \bibfield  {author} {\bibinfo {author} {\bibfnamefont {C.~V.~R.}\
  \bibnamefont {Board}}\ and\ \bibinfo {author} {\bibfnamefont {J.~D.}\
  \bibnamefont {Barrow}},\ }\bibfield  {title} {\enquote {\bibinfo {title}
  {{Cosmological Models in Energy-Momentum-Squared Gravity}},}\ }\href
  {\doibase 10.1103/PhysRevD.96.123517} {\bibfield  {journal} {\bibinfo
  {journal} {Phys. Rev.}\ }\textbf {\bibinfo {volume} {D 96}},\ \bibinfo
  {pages} {123517} (\bibinfo {year} {2017})},\ \Eprint
  {http://arxiv.org/abs/1709.09501} {arXiv:1709.09501 [gr-qc]} \BibitemShut
  {NoStop}%
%%CITATION = ARXIV:1709.09501;%%
\bibitem [{\citenamefont {{Ashtekar}}\ and\ \citenamefont
  {{Singh}}(2011)}]{LQC11}%
  \BibitemOpen
  \bibfield  {author} {\bibinfo {author} {\bibfnamefont {A.}~\bibnamefont
  {{Ashtekar}}}\ and\ \bibinfo {author} {\bibfnamefont {P.}~\bibnamefont
  {{Singh}}},\ }\bibfield  {title} {\enquote {\bibinfo {title} {{Loop quantum
  cosmology: a status report}},}\ }\href {\doibase
  10.1088/0264-9381/28/21/213001} {\bibfield  {journal} {\bibinfo  {journal}
  {Classical and Quantum Gravity}\ }\textbf {\bibinfo {volume} {28}},\ \bibinfo
  {eid} {213001} (\bibinfo {year} {2011})},\ \Eprint
  {http://arxiv.org/abs/1108.0893} {arXiv:1108.0893 [gr-qc]} \BibitemShut
  {NoStop}%
\bibitem [{\citenamefont {{Brax}}\ and\ \citenamefont {{van de
  Bruck}}(2003)}]{BraneW1}%
  \BibitemOpen
  \bibfield  {author} {\bibinfo {author} {\bibfnamefont {P.}~\bibnamefont
  {{Brax}}}\ and\ \bibinfo {author} {\bibfnamefont {C.}~\bibnamefont {{van de
  Bruck}}},\ }\bibfield  {title} {\enquote {\bibinfo {title} {{TOPICAL REVIEW:
  Cosmology and brane worlds: a review}},}\ }\href {\doibase
  10.1088/0264-9381/20/9/202} {\bibfield  {journal} {\bibinfo  {journal}
  {Classical and Quantum Gravity}\ }\textbf {\bibinfo {volume} {20}},\ \bibinfo
  {pages} {R201--R232} (\bibinfo {year} {2003})},\ \Eprint
  {http://arxiv.org/abs/hep-th/0303095} {hep-th/0303095} \BibitemShut {NoStop}%
\bibitem [{\citenamefont {{Ambartsumyan}}\ and\ \citenamefont
  {{Saakyan}}(1960)}]{amb60}%
  \BibitemOpen
  \bibfield  {author} {\bibinfo {author} {\bibfnamefont {V.~A.}\ \bibnamefont
  {{Ambartsumyan}}}\ and\ \bibinfo {author} {\bibfnamefont {G.~S.}\
  \bibnamefont {{Saakyan}}},\ }\bibfield  {title} {\enquote {\bibinfo {title}
  {{The Degenerate Superdense Gas of Elementary Particles}},}\ }\href@noop {}
  {\bibfield  {journal} {\bibinfo  {journal} {Soviet Astronomy}\ }\textbf
  {\bibinfo {volume} {4}},\ \bibinfo {pages} {187} (\bibinfo {year}
  {1960})}\BibitemShut {NoStop}%
\bibitem [{\citenamefont {{Chatterjee}}\ and\ \citenamefont
  {{Vida{\~n}a}}(2016)}]{cha16}%
  \BibitemOpen
  \bibfield  {author} {\bibinfo {author} {\bibfnamefont {D.}~\bibnamefont
  {{Chatterjee}}}\ and\ \bibinfo {author} {\bibfnamefont {I.}~\bibnamefont
  {{Vida{\~n}a}}},\ }\bibfield  {title} {\enquote {\bibinfo {title} {{Do
  hyperons exist in the interior of neutron stars?}}}\ }\href {\doibase
  10.1140/epja/i2016-16029-x} {\bibfield  {journal} {\bibinfo  {journal}
  {European Physical Journal A}\ }\textbf {\bibinfo {volume} {52}},\ \bibinfo
  {eid} {29} (\bibinfo {year} {2016})},\ \Eprint
  {http://arxiv.org/abs/1510.06306} {arXiv:1510.06306 [nucl-th]} \BibitemShut
  {NoStop}%
\bibitem [{\citenamefont {{Demorest}}\ \emph {et~al.}(2010)\citenamefont
  {{Demorest}}, \citenamefont {{Pennucci}}, \citenamefont {{Ransom}},
  \citenamefont {{Roberts}},\ and\ \citenamefont {{Hessels}}}]{dem+10}%
  \BibitemOpen
  \bibfield  {author} {\bibinfo {author} {\bibfnamefont {P.~B.}\ \bibnamefont
  {{Demorest}}}, \bibinfo {author} {\bibfnamefont {T.}~\bibnamefont
  {{Pennucci}}}, \bibinfo {author} {\bibfnamefont {S.~M.}\ \bibnamefont
  {{Ransom}}}, \bibinfo {author} {\bibfnamefont {M.~S.~E.}\ \bibnamefont
  {{Roberts}}}, \ and\ \bibinfo {author} {\bibfnamefont {J.~W.~T.}\
  \bibnamefont {{Hessels}}},\ }\bibfield  {title} {\enquote {\bibinfo {title}
  {{A two-solar-mass neutron star measured using Shapiro delay}},}\ }\href
  {\doibase 10.1038/nature09466} {\bibfield  {journal} {\bibinfo  {journal}
  {\nat}\ }\textbf {\bibinfo {volume} {467}},\ \bibinfo {pages} {1081--1083}
  (\bibinfo {year} {2010})},\ \Eprint {http://arxiv.org/abs/1010.5788}
  {arXiv:1010.5788 [astro-ph.HE]} \BibitemShut {NoStop}%
\bibitem [{\citenamefont {{Antoniadis}}\ \emph {et~al.}(2013)\citenamefont
  {{Antoniadis}}, \citenamefont {{Freire}}, \citenamefont {{Wex}},
  \citenamefont {{Tauris}}, \citenamefont {{Lynch}}, \citenamefont {{van
  Kerkwijk}}, \citenamefont {{Kramer}}, \citenamefont {{Bassa}}, \citenamefont
  {{Dhillon}}, \citenamefont {{Driebe}}, \citenamefont {{Hessels}},
  \citenamefont {{Kaspi}}, \citenamefont {{Kondratiev}}, \citenamefont
  {{Langer}}, \citenamefont {{Marsh}}, \citenamefont {{McLaughlin}},
  \citenamefont {{Pennucci}}, \citenamefont {{Ransom}}, \citenamefont
  {{Stairs}}, \citenamefont {{van Leeuwen}}, \citenamefont {{Verbiest}},\ and\
  \citenamefont {{Whelan}}}]{ant+13}%
  \BibitemOpen
  \bibfield  {author} {\bibinfo {author} {\bibfnamefont {J.}~\bibnamefont
  {{Antoniadis}}}, \bibinfo {author} {\bibfnamefont {P.~C.~C.}\ \bibnamefont
  {{Freire}}}, \bibinfo {author} {\bibfnamefont {N.}~\bibnamefont {{Wex}}},
  \bibinfo {author} {\bibfnamefont {T.~M.}\ \bibnamefont {{Tauris}}}, \bibinfo
  {author} {\bibfnamefont {R.~S.}\ \bibnamefont {{Lynch}}}, \bibinfo {author}
  {\bibfnamefont {M.~H.}\ \bibnamefont {{van Kerkwijk}}}, \bibinfo {author}
  {\bibfnamefont {M.}~\bibnamefont {{Kramer}}}, \bibinfo {author}
  {\bibfnamefont {C.}~\bibnamefont {{Bassa}}}, \bibinfo {author} {\bibfnamefont
  {V.~S.}\ \bibnamefont {{Dhillon}}}, \bibinfo {author} {\bibfnamefont
  {T.}~\bibnamefont {{Driebe}}}, \bibinfo {author} {\bibfnamefont {J.~W.~T.}\
  \bibnamefont {{Hessels}}}, \bibinfo {author} {\bibfnamefont {V.~M.}\
  \bibnamefont {{Kaspi}}}, \bibinfo {author} {\bibfnamefont {V.~I.}\
  \bibnamefont {{Kondratiev}}}, \bibinfo {author} {\bibfnamefont
  {N.}~\bibnamefont {{Langer}}}, \bibinfo {author} {\bibfnamefont {T.~R.}\
  \bibnamefont {{Marsh}}}, \bibinfo {author} {\bibfnamefont {M.~A.}\
  \bibnamefont {{McLaughlin}}}, \bibinfo {author} {\bibfnamefont {T.~T.}\
  \bibnamefont {{Pennucci}}}, \bibinfo {author} {\bibfnamefont {S.~M.}\
  \bibnamefont {{Ransom}}}, \bibinfo {author} {\bibfnamefont {I.~H.}\
  \bibnamefont {{Stairs}}}, \bibinfo {author} {\bibfnamefont {J.}~\bibnamefont
  {{van Leeuwen}}}, \bibinfo {author} {\bibfnamefont {J.~P.~W.}\ \bibnamefont
  {{Verbiest}}}, \ and\ \bibinfo {author} {\bibfnamefont {D.~G.}\ \bibnamefont
  {{Whelan}}},\ }\bibfield  {title} {\enquote {\bibinfo {title} {{A Massive
  Pulsar in a Compact Relativistic Binary}},}\ }\href {\doibase
  10.1126/science.1233232} {\bibfield  {journal} {\bibinfo  {journal}
  {Science}\ }\textbf {\bibinfo {volume} {340}},\ \bibinfo {pages} {1233232}
  (\bibinfo {year} {2013})},\ \Eprint {http://arxiv.org/abs/1304.6875}
  {arXiv:1304.6875 [astro-ph.HE]} \BibitemShut {NoStop}%
\bibitem [{\citenamefont {{Astashenok}}\ \emph {et~al.}(2014)\citenamefont
  {{Astashenok}}, \citenamefont {{Capozziello}},\ and\ \citenamefont
  {{Odintsov}}}]{ast+14}%
  \BibitemOpen
  \bibfield  {author} {\bibinfo {author} {\bibfnamefont {A.~V.}\ \bibnamefont
  {{Astashenok}}}, \bibinfo {author} {\bibfnamefont {S.}~\bibnamefont
  {{Capozziello}}}, \ and\ \bibinfo {author} {\bibfnamefont {S.~D.}\
  \bibnamefont {{Odintsov}}},\ }\bibfield  {title} {\enquote {\bibinfo {title}
  {{Maximal neutron star mass and the resolution of the hyperon puzzle in
  modified gravity}},}\ }\href {\doibase 10.1103/PhysRevD.89.103509} {\bibfield
   {journal} {\bibinfo  {journal} {\prd}\ }\textbf {\bibinfo {volume} {89}},\
  \bibinfo {eid} {103509} (\bibinfo {year} {2014})},\ \Eprint
  {http://arxiv.org/abs/1401.4546} {arXiv:1401.4546 [gr-qc]} \BibitemShut
  {NoStop}%
\bibitem [{\citenamefont {{Germani}}\ and\ \citenamefont
  {{Maartens}}(2001)}]{Maartens01}%
  \BibitemOpen
  \bibfield  {author} {\bibinfo {author} {\bibfnamefont {C.}~\bibnamefont
  {{Germani}}}\ and\ \bibinfo {author} {\bibfnamefont {R.}~\bibnamefont
  {{Maartens}}},\ }\bibfield  {title} {\enquote {\bibinfo {title} {{Stars in
  the braneworld}},}\ }\href {\doibase 10.1103/PhysRevD.64.124010} {\bibfield
  {journal} {\bibinfo  {journal} {\prd}\ }\textbf {\bibinfo {volume} {64}},\
  \bibinfo {pages} {124010} (\bibinfo {year} {2001})},\ \Eprint
  {http://arxiv.org/abs/hep-th/0107011} {hep-th/0107011} \BibitemShut {NoStop}%
\bibitem [{\citenamefont {{Lugones}}\ and\ \citenamefont
  {{Arba{\~n}il}}(2017)}]{Lugones17}%
  \BibitemOpen
  \bibfield  {author} {\bibinfo {author} {\bibfnamefont {G.}~\bibnamefont
  {{Lugones}}}\ and\ \bibinfo {author} {\bibfnamefont {J.~D.~V.}\ \bibnamefont
  {{Arba{\~n}il}}},\ }\bibfield  {title} {\enquote {\bibinfo {title} {{Compact
  stars in the braneworld: A new branch of stellar configurations with
  arbitrarily large mass}},}\ }\href {\doibase 10.1103/PhysRevD.95.064022}
  {\bibfield  {journal} {\bibinfo  {journal} {\prd}\ }\textbf {\bibinfo
  {volume} {95}},\ \bibinfo {eid} {064022} (\bibinfo {year} {2017})},\ \Eprint
  {http://arxiv.org/abs/1702.07824} {arXiv:1702.07824 [gr-qc]} \BibitemShut
  {NoStop}%
\bibitem [{\citenamefont {{Prasetyo}}\ \emph {et~al.}(2018)\citenamefont
  {{Prasetyo}}, \citenamefont {{Husin}}, \citenamefont {{Qauli}}, \citenamefont
  {{Ramadhan}},\ and\ \citenamefont {{Sulaksono}}}]{Prasetyo18}%
  \BibitemOpen
  \bibfield  {author} {\bibinfo {author} {\bibfnamefont {I.}~\bibnamefont
  {{Prasetyo}}}, \bibinfo {author} {\bibfnamefont {I.}~\bibnamefont {{Husin}}},
  \bibinfo {author} {\bibfnamefont {A.~I.}\ \bibnamefont {{Qauli}}}, \bibinfo
  {author} {\bibfnamefont {H.~S.}\ \bibnamefont {{Ramadhan}}}, \ and\ \bibinfo
  {author} {\bibfnamefont {A.}~\bibnamefont {{Sulaksono}}},\ }\bibfield
  {title} {\enquote {\bibinfo {title} {{Neutron stars in the braneworld within
  the Eddington-inspired Born-Infeld gravity}},}\ }\href {\doibase
  10.1088/1475-7516/2018/01/027} {\bibfield  {journal} {\bibinfo  {journal}
  {\jcap}\ }\textbf {\bibinfo {volume} {1}},\ \bibinfo {eid} {027} (\bibinfo
  {year} {2018})},\ \Eprint {http://arxiv.org/abs/1708.04837}
  {arXiv:1708.04837} \BibitemShut {NoStop}%
\bibitem [{\citenamefont {{Arapo{\u g}lu}}\ \emph {et~al.}(2011)\citenamefont
  {{Arapo{\u g}lu}}, \citenamefont {{Deliduman}},\ and\ \citenamefont {{Ek{\c
  s}i}}}]{ara+11}%
  \BibitemOpen
  \bibfield  {author} {\bibinfo {author} {\bibfnamefont {S.}~\bibnamefont
  {{Arapo{\u g}lu}}}, \bibinfo {author} {\bibfnamefont {C.}~\bibnamefont
  {{Deliduman}}}, \ and\ \bibinfo {author} {\bibfnamefont {K.~Y.}\ \bibnamefont
  {{Ek{\c s}i}}},\ }\bibfield  {title} {\enquote {\bibinfo {title}
  {{Constraints on perturbative f(R) gravity via neutron stars}},}\ }\href
  {\doibase 10.1088/1475-7516/2011/07/020} {\bibfield  {journal} {\bibinfo
  {journal} {\jcap}\ }\textbf {\bibinfo {volume} {7}},\ \bibinfo {eid} {020}
  (\bibinfo {year} {2011})},\ \Eprint {http://arxiv.org/abs/1003.3179}
  {arXiv:1003.3179 [gr-qc]} \BibitemShut {NoStop}%
\bibitem [{\citenamefont {{Ek{\c s}i}}\ \emph {et~al.}(2014)\citenamefont
  {{Ek{\c s}i}}, \citenamefont {{G{\"u}ng{\"o}r}},\ and\ \citenamefont
  {{T{\"u}rko{\v g}lu}}}]{eks+14}%
  \BibitemOpen
  \bibfield  {author} {\bibinfo {author} {\bibfnamefont {K.~Y.}\ \bibnamefont
  {{Ek{\c s}i}}}, \bibinfo {author} {\bibfnamefont {C.}~\bibnamefont
  {{G{\"u}ng{\"o}r}}}, \ and\ \bibinfo {author} {\bibfnamefont {M.~M.}\
  \bibnamefont {{T{\"u}rko{\v g}lu}}},\ }\bibfield  {title} {\enquote {\bibinfo
  {title} {{What does a measurement of mass and/or radius of a neutron star
  constrain: Equation of state or gravity?}}}\ }\href {\doibase
  10.1103/PhysRevD.89.063003} {\bibfield  {journal} {\bibinfo  {journal}
  {\prd}\ }\textbf {\bibinfo {volume} {89}},\ \bibinfo {eid} {063003} (\bibinfo
  {year} {2014})},\ \Eprint {http://arxiv.org/abs/1402.0488} {arXiv:1402.0488
  [astro-ph.HE]} \BibitemShut {NoStop}%
\bibitem [{\citenamefont {Hawking}(1972)}]{Haw72}%
  \BibitemOpen
  \bibfield  {author} {\bibinfo {author} {\bibfnamefont {S.~W.}\ \bibnamefont
  {Hawking}},\ }\bibfield  {title} {\enquote {\bibinfo {title} {Black holes in
  the brans-dicke},}\ }\href {\doibase 10.1007/BF01877518} {\bibfield
  {journal} {\bibinfo  {journal} {Communications in Mathematical Physics}\
  }\textbf {\bibinfo {volume} {25}},\ \bibinfo {pages} {167--171} (\bibinfo
  {year} {1972})}\BibitemShut {NoStop}%
\bibitem [{\citenamefont {Bekenstein}(1972)}]{bek72}%
  \BibitemOpen
  \bibfield  {author} {\bibinfo {author} {\bibfnamefont {J.~D.}\ \bibnamefont
  {Bekenstein}},\ }\bibfield  {title} {\enquote {\bibinfo {title} {Nonexistence
  of baryon number for black holes. ii},}\ }\href {\doibase
  10.1103/PhysRevD.5.2403} {\bibfield  {journal} {\bibinfo  {journal} {Phys.
  Rev. D}\ }\textbf {\bibinfo {volume} {5}},\ \bibinfo {pages} {2403--2412}
  (\bibinfo {year} {1972})}\BibitemShut {NoStop}%
\bibitem [{\citenamefont {Bekenstein}\ and\ \citenamefont
  {Meisels}(1978)}]{bek+78}%
  \BibitemOpen
  \bibfield  {author} {\bibinfo {author} {\bibfnamefont {J.~D.}\ \bibnamefont
  {Bekenstein}}\ and\ \bibinfo {author} {\bibfnamefont {A.}~\bibnamefont
  {Meisels}},\ }\bibfield  {title} {\enquote {\bibinfo {title} {General
  relativity without general relativity},}\ }\href {\doibase
  10.1103/PhysRevD.18.4378} {\bibfield  {journal} {\bibinfo  {journal} {Phys.
  Rev. D}\ }\textbf {\bibinfo {volume} {18}},\ \bibinfo {pages} {4378--4386}
  (\bibinfo {year} {1978})}\BibitemShut {NoStop}%
\bibitem [{\citenamefont {Sotiriou}\ and\ \citenamefont
  {Faraoni}(2012)}]{sot+12}%
  \BibitemOpen
  \bibfield  {author} {\bibinfo {author} {\bibfnamefont {T.~P.}\ \bibnamefont
  {Sotiriou}}\ and\ \bibinfo {author} {\bibfnamefont {V.}~\bibnamefont
  {Faraoni}},\ }\bibfield  {title} {\enquote {\bibinfo {title} {Black holes in
  scalar-tensor gravity},}\ }\href {\doibase 10.1103/PhysRevLett.108.081103}
  {\bibfield  {journal} {\bibinfo  {journal} {Phys. Rev. Lett.}\ }\textbf
  {\bibinfo {volume} {108}},\ \bibinfo {pages} {081103} (\bibinfo {year}
  {2012})}\BibitemShut {NoStop}%
\bibitem [{\citenamefont {{Psaltis}}\ \emph {et~al.}(2008)\citenamefont
  {{Psaltis}}, \citenamefont {{Perrodin}}, \citenamefont {{Dienes}},\ and\
  \citenamefont {{Mocioiu}}}]{psa+08}%
  \BibitemOpen
  \bibfield  {author} {\bibinfo {author} {\bibfnamefont {D.}~\bibnamefont
  {{Psaltis}}}, \bibinfo {author} {\bibfnamefont {D.}~\bibnamefont
  {{Perrodin}}}, \bibinfo {author} {\bibfnamefont {K.~R.}\ \bibnamefont
  {{Dienes}}}, \ and\ \bibinfo {author} {\bibfnamefont {I.}~\bibnamefont
  {{Mocioiu}}},\ }\bibfield  {title} {\enquote {\bibinfo {title} {{Kerr Black
  Holes Are Not Unique to General Relativity}},}\ }\href {\doibase
  10.1103/PhysRevLett.100.091101} {\bibfield  {journal} {\bibinfo  {journal}
  {Physical Review Letters}\ }\textbf {\bibinfo {volume} {100}},\ \bibinfo
  {eid} {091101} (\bibinfo {year} {2008})}\BibitemShut {NoStop}%
\bibitem [{\citenamefont {{DeDeo}}\ and\ \citenamefont
  {{Psaltis}}(2003)}]{ded03}%
  \BibitemOpen
  \bibfield  {author} {\bibinfo {author} {\bibfnamefont {S.}~\bibnamefont
  {{DeDeo}}}\ and\ \bibinfo {author} {\bibfnamefont {D.}~\bibnamefont
  {{Psaltis}}},\ }\bibfield  {title} {\enquote {\bibinfo {title} {{Towards New
  Tests of Strong-Field Gravity with Measurements of Surface Atomic Line
  Redshifts from Neutron Stars}},}\ }\href {\doibase
  10.1103/PhysRevLett.90.141101} {\bibfield  {journal} {\bibinfo  {journal}
  {Physical Review Letters}\ }\textbf {\bibinfo {volume} {90}},\ \bibinfo {eid}
  {141101} (\bibinfo {year} {2003})},\ \Eprint
  {http://arxiv.org/abs/astro-ph/0302095} {astro-ph/0302095} \BibitemShut
  {NoStop}%
\bibitem [{\citenamefont {{Cooney}}\ \emph {et~al.}(2010)\citenamefont
  {{Cooney}}, \citenamefont {{Dedeo}},\ and\ \citenamefont
  {{Psaltis}}}]{coo+10}%
  \BibitemOpen
  \bibfield  {author} {\bibinfo {author} {\bibfnamefont {A.}~\bibnamefont
  {{Cooney}}}, \bibinfo {author} {\bibfnamefont {S.}~\bibnamefont {{Dedeo}}}, \
  and\ \bibinfo {author} {\bibfnamefont {D.}~\bibnamefont {{Psaltis}}},\
  }\bibfield  {title} {\enquote {\bibinfo {title} {{Neutron stars in f(R)
  gravity with perturbative constraints}},}\ }\href {\doibase
  10.1103/PhysRevD.82.064033} {\bibfield  {journal} {\bibinfo  {journal}
  {\prd}\ }\textbf {\bibinfo {volume} {82}},\ \bibinfo {eid} {064033} (\bibinfo
  {year} {2010})},\ \Eprint {http://arxiv.org/abs/0910.5480} {arXiv:0910.5480
  [astro-ph.HE]} \BibitemShut {NoStop}%
\bibitem [{\citenamefont {{Capozziello}}\ \emph {et~al.}(2011)\citenamefont
  {{Capozziello}}, \citenamefont {{de Laurentis}}, \citenamefont {{Odintsov}},\
  and\ \citenamefont {{Stabile}}}]{cap+11}%
  \BibitemOpen
  \bibfield  {author} {\bibinfo {author} {\bibfnamefont {S.}~\bibnamefont
  {{Capozziello}}}, \bibinfo {author} {\bibfnamefont {M.}~\bibnamefont {{de
  Laurentis}}}, \bibinfo {author} {\bibfnamefont {S.~D.}\ \bibnamefont
  {{Odintsov}}}, \ and\ \bibinfo {author} {\bibfnamefont {A.}~\bibnamefont
  {{Stabile}}},\ }\bibfield  {title} {\enquote {\bibinfo {title} {{Hydrostatic
  equilibrium and stellar structure in f(R) gravity}},}\ }\href {\doibase
  10.1103/PhysRevD.83.064004} {\bibfield  {journal} {\bibinfo  {journal}
  {\prd}\ }\textbf {\bibinfo {volume} {83}},\ \bibinfo {eid} {064004} (\bibinfo
  {year} {2011})},\ \Eprint {http://arxiv.org/abs/1101.0219} {arXiv:1101.0219
  [gr-qc]} \BibitemShut {NoStop}%
\bibitem [{\citenamefont {{Pani}}\ \emph {et~al.}(2011)\citenamefont {{Pani}},
  \citenamefont {{Cardoso}},\ and\ \citenamefont {{Delsate}}}]{pan+11}%
  \BibitemOpen
  \bibfield  {author} {\bibinfo {author} {\bibfnamefont {P.}~\bibnamefont
  {{Pani}}}, \bibinfo {author} {\bibfnamefont {V.}~\bibnamefont {{Cardoso}}}, \
  and\ \bibinfo {author} {\bibfnamefont {T.}~\bibnamefont {{Delsate}}},\
  }\bibfield  {title} {\enquote {\bibinfo {title} {{Compact Stars in Eddington
  Inspired Gravity}},}\ }\href {\doibase 10.1103/PhysRevLett.107.031101}
  {\bibfield  {journal} {\bibinfo  {journal} {Physical Review Letters}\
  }\textbf {\bibinfo {volume} {107}},\ \bibinfo {eid} {031101} (\bibinfo {year}
  {2011})},\ \Eprint {http://arxiv.org/abs/1106.3569} {arXiv:1106.3569 [gr-qc]}
  \BibitemShut {NoStop}%
\bibitem [{\citenamefont {{Deliduman}}\ \emph {et~al.}(2012)\citenamefont
  {{Deliduman}}, \citenamefont {{Ek{\c s}i}},\ and\ \citenamefont {{Kele{\c
  s}}}}]{del+12}%
  \BibitemOpen
  \bibfield  {author} {\bibinfo {author} {\bibfnamefont {C.}~\bibnamefont
  {{Deliduman}}}, \bibinfo {author} {\bibfnamefont {K.~Y.}\ \bibnamefont
  {{Ek{\c s}i}}}, \ and\ \bibinfo {author} {\bibfnamefont {V.}~\bibnamefont
  {{Kele{\c s}}}},\ }\bibfield  {title} {\enquote {\bibinfo {title} {{Neutron
  star solutions in perturbative quadratic gravity}},}\ }\href {\doibase
  10.1088/1475-7516/2012/05/036} {\bibfield  {journal} {\bibinfo  {journal}
  {\jcap}\ }\textbf {\bibinfo {volume} {5}},\ \bibinfo {eid} {036} (\bibinfo
  {year} {2012})},\ \Eprint {http://arxiv.org/abs/1112.4154} {arXiv:1112.4154
  [gr-qc]} \BibitemShut {NoStop}%
\bibitem [{\citenamefont {{Astashenok}}\ \emph {et~al.}(2013)\citenamefont
  {{Astashenok}}, \citenamefont {{Capozziello}},\ and\ \citenamefont
  {{Odintsov}}}]{ast+13}%
  \BibitemOpen
  \bibfield  {author} {\bibinfo {author} {\bibfnamefont {A.~V.}\ \bibnamefont
  {{Astashenok}}}, \bibinfo {author} {\bibfnamefont {S.}~\bibnamefont
  {{Capozziello}}}, \ and\ \bibinfo {author} {\bibfnamefont {S.~D.}\
  \bibnamefont {{Odintsov}}},\ }\bibfield  {title} {\enquote {\bibinfo {title}
  {{Further stable neutron star models from f(R) gravity}},}\ }\href {\doibase
  10.1088/1475-7516/2013/12/040} {\bibfield  {journal} {\bibinfo  {journal}
  {\jcap}\ }\textbf {\bibinfo {volume} {12}},\ \bibinfo {eid} {040} (\bibinfo
  {year} {2013})},\ \Eprint {http://arxiv.org/abs/1309.1978} {arXiv:1309.1978
  [gr-qc]} \BibitemShut {NoStop}%
\bibitem [{\citenamefont {{Yazadjiev}}\ \emph {et~al.}(2014a)\citenamefont
  {{Yazadjiev}}, \citenamefont {{Doneva}}, \citenamefont {{Kokkotas}},\ and\
  \citenamefont {{Staykov}}}]{yaz+14}%
  \BibitemOpen
  \bibfield  {author} {\bibinfo {author} {\bibfnamefont {S.~S.}\ \bibnamefont
  {{Yazadjiev}}}, \bibinfo {author} {\bibfnamefont {D.~D.}\ \bibnamefont
  {{Doneva}}}, \bibinfo {author} {\bibfnamefont {K.~D.}\ \bibnamefont
  {{Kokkotas}}}, \ and\ \bibinfo {author} {\bibfnamefont {K.~V.}\ \bibnamefont
  {{Staykov}}},\ }\bibfield  {title} {\enquote {\bibinfo {title}
  {{Non-perturbative and self-consistent models of neutron stars in R-squared
  gravity}},}\ }\href {\doibase 10.1088/1475-7516/2014/06/003} {\bibfield
  {journal} {\bibinfo  {journal} {\jcap}\ }\textbf {\bibinfo {volume} {6}},\
  \bibinfo {eid} {003} (\bibinfo {year} {2014a})},\ \Eprint
  {http://arxiv.org/abs/1402.4469} {arXiv:1402.4469 [gr-qc]} \BibitemShut
  {NoStop}%
\bibitem [{\citenamefont {{Ganguly}}\ \emph {et~al.}(2014)\citenamefont
  {{Ganguly}}, \citenamefont {{Gannouji}}, \citenamefont {{Goswami}},\ and\
  \citenamefont {{Ray}}}]{gan+14}%
  \BibitemOpen
  \bibfield  {author} {\bibinfo {author} {\bibfnamefont {A.}~\bibnamefont
  {{Ganguly}}}, \bibinfo {author} {\bibfnamefont {R.}~\bibnamefont
  {{Gannouji}}}, \bibinfo {author} {\bibfnamefont {R.}~\bibnamefont
  {{Goswami}}}, \ and\ \bibinfo {author} {\bibfnamefont {S.}~\bibnamefont
  {{Ray}}},\ }\bibfield  {title} {\enquote {\bibinfo {title} {{Neutron stars in
  the Starobinsky model}},}\ }\href {\doibase 10.1103/PhysRevD.89.064019}
  {\bibfield  {journal} {\bibinfo  {journal} {Phys. Rev. D.}\ }\textbf
  {\bibinfo {volume} {89}},\ \bibinfo {eid} {064019} (\bibinfo {year}
  {2014})},\ \Eprint {http://arxiv.org/abs/1309.3279} {arXiv:1309.3279 [gr-qc]}
  \BibitemShut {NoStop}%
\bibitem [{\citenamefont {{Astashenok}}\ \emph {et~al.}(2015a)\citenamefont
  {{Astashenok}}, \citenamefont {{Capozziello}},\ and\ \citenamefont
  {{Odintsov}}}]{ast+15a}%
  \BibitemOpen
  \bibfield  {author} {\bibinfo {author} {\bibfnamefont {A.~V.}\ \bibnamefont
  {{Astashenok}}}, \bibinfo {author} {\bibfnamefont {S.}~\bibnamefont
  {{Capozziello}}}, \ and\ \bibinfo {author} {\bibfnamefont {S.~D.}\
  \bibnamefont {{Odintsov}}},\ }\bibfield  {title} {\enquote {\bibinfo {title}
  {{Extreme neutron stars from Extended Theories of Gravity}},}\ }\href
  {\doibase 10.1088/1475-7516/2015/01/001} {\bibfield  {journal} {\bibinfo
  {journal} {\jcap}\ }\textbf {\bibinfo {volume} {1}},\ \bibinfo {eid} {001}
  (\bibinfo {year} {2015a})},\ \Eprint {http://arxiv.org/abs/1408.3856}
  {arXiv:1408.3856 [gr-qc]} \BibitemShut {NoStop}%
\bibitem [{\citenamefont {{Astashenok}}\ \emph {et~al.}(2015b)\citenamefont
  {{Astashenok}}, \citenamefont {{Capozziello}},\ and\ \citenamefont
  {{Odintsov}}}]{ast+15b}%
  \BibitemOpen
  \bibfield  {author} {\bibinfo {author} {\bibfnamefont {A.~V.}\ \bibnamefont
  {{Astashenok}}}, \bibinfo {author} {\bibfnamefont {S.}~\bibnamefont
  {{Capozziello}}}, \ and\ \bibinfo {author} {\bibfnamefont {S.~D.}\
  \bibnamefont {{Odintsov}}},\ }\bibfield  {title} {\enquote {\bibinfo {title}
  {{Magnetic neutron stars in f(R) gravity}},}\ }\href {\doibase
  10.1007/s10509-014-2182-6} {\bibfield  {journal} {\bibinfo  {journal}
  {\apss}\ }\textbf {\bibinfo {volume} {355}},\ \bibinfo {pages} {333--341}
  (\bibinfo {year} {2015b})},\ \Eprint {http://arxiv.org/abs/1405.6663}
  {arXiv:1405.6663 [gr-qc]} \BibitemShut {NoStop}%
\bibitem [{\citenamefont {{Capozziello}}\ \emph {et~al.}(2016)\citenamefont
  {{Capozziello}}, \citenamefont {{De Laurentis}}, \citenamefont
  {{Farinelli}},\ and\ \citenamefont {{Odintsov}}}]{cap+16}%
  \BibitemOpen
  \bibfield  {author} {\bibinfo {author} {\bibfnamefont {S.}~\bibnamefont
  {{Capozziello}}}, \bibinfo {author} {\bibfnamefont {M.}~\bibnamefont {{De
  Laurentis}}}, \bibinfo {author} {\bibfnamefont {R.}~\bibnamefont
  {{Farinelli}}}, \ and\ \bibinfo {author} {\bibfnamefont {S.~D.}\ \bibnamefont
  {{Odintsov}}},\ }\bibfield  {title} {\enquote {\bibinfo {title} {{Mass-radius
  relation for neutron stars in f(R) gravity}},}\ }\href {\doibase
  10.1103/PhysRevD.93.023501} {\bibfield  {journal} {\bibinfo  {journal}
  {\prd}\ }\textbf {\bibinfo {volume} {93}},\ \bibinfo {eid} {023501} (\bibinfo
  {year} {2016})},\ \Eprint {http://arxiv.org/abs/1509.04163} {arXiv:1509.04163
  [gr-qc]} \BibitemShut {NoStop}%
\bibitem [{\citenamefont {{Arapo{\u g}lu}}\ \emph {et~al.}(2017)\citenamefont
  {{Arapo{\u g}lu}}, \citenamefont {{{\c C}{\i}k{\i}nto{\u g}lu}},\ and\
  \citenamefont {{Ek{\c s}i}}}]{ara+17}%
  \BibitemOpen
  \bibfield  {author} {\bibinfo {author} {\bibfnamefont {S.}~\bibnamefont
  {{Arapo{\u g}lu}}}, \bibinfo {author} {\bibfnamefont {S.}~\bibnamefont {{{\c
  C}{\i}k{\i}nto{\u g}lu}}}, \ and\ \bibinfo {author} {\bibfnamefont {K.~Y.}\
  \bibnamefont {{Ek{\c s}i}}},\ }\bibfield  {title} {\enquote {\bibinfo {title}
  {Relativistic stars in starobinsky gravity with the matched asymptotic
  expansions method},}\ }\href {\doibase 10.1103/PhysRevD.96.084040} {\bibfield
   {journal} {\bibinfo  {journal} {Phys. Rev. D}\ }\textbf {\bibinfo {volume}
  {96}},\ \bibinfo {pages} {084040} (\bibinfo {year} {2017})}\BibitemShut
  {NoStop}%
\bibitem [{\citenamefont {{Astashenok}}\ \emph {et~al.}(2017)\citenamefont
  {{Astashenok}}, \citenamefont {{de la Cruz-Dombriz}},\ and\ \citenamefont
  {{Odintsov}}}]{ast+17}%
  \BibitemOpen
  \bibfield  {author} {\bibinfo {author} {\bibfnamefont {A.~V.}\ \bibnamefont
  {{Astashenok}}}, \bibinfo {author} {\bibfnamefont {A.}~\bibnamefont {{de la
  Cruz-Dombriz}}}, \ and\ \bibinfo {author} {\bibfnamefont {S.~D.}\
  \bibnamefont {{Odintsov}}},\ }\bibfield  {title} {\enquote {\bibinfo {title}
  {{The realistic models of relativistic stars in $f(R)=R+\alpha R^{2}$
  gravity}},}\ }\href {\doibase 10.1088/1361-6382/aa8971} {\bibfield  {journal}
  {\bibinfo  {journal} {Class. Quantum Grav.}\ }\textbf {\bibinfo {volume}
  {34}},\ \bibinfo {eid} {205008} (\bibinfo {year} {2017})},\ \Eprint
  {http://arxiv.org/abs/1704.08311} {arXiv:1704.08311 [gr-qc]} \BibitemShut
  {NoStop}%
\bibitem [{\citenamefont {{Doneva}}\ and\ \citenamefont
  {{Pappas}}(2017)}]{don17}%
  \BibitemOpen
  \bibfield  {author} {\bibinfo {author} {\bibfnamefont {D.~D.}\ \bibnamefont
  {{Doneva}}}\ and\ \bibinfo {author} {\bibfnamefont {G.}~\bibnamefont
  {{Pappas}}},\ }\bibfield  {title} {\enquote {\bibinfo {title} {{Universal
  Relations and Alternative Gravity Theories}},}\ }\href@noop {} {\bibfield
  {journal} {\bibinfo  {journal} {ArXiv e-prints}\ } (\bibinfo {year}
  {2017})},\ \Eprint {http://arxiv.org/abs/1709.08046} {arXiv:1709.08046
  [gr-qc]} \BibitemShut {NoStop}%
\bibitem [{\citenamefont {{Bertolami}}\ \emph {et~al.}(2008)\citenamefont
  {{Bertolami}}, \citenamefont {{Lobo}},\ and\ \citenamefont
  {{P{\'a}ramos}}}]{Bertolami:2008ab}%
  \BibitemOpen
  \bibfield  {author} {\bibinfo {author} {\bibfnamefont {O.}~\bibnamefont
  {{Bertolami}}}, \bibinfo {author} {\bibfnamefont {F.~S.~N.}\ \bibnamefont
  {{Lobo}}}, \ and\ \bibinfo {author} {\bibfnamefont {J.}~\bibnamefont
  {{P{\'a}ramos}}},\ }\bibfield  {title} {\enquote {\bibinfo {title}
  {{Nonminimal coupling of perfect fluids to curvature}},}\ }\href {\doibase
  10.1103/PhysRevD.78.064036} {\bibfield  {journal} {\bibinfo  {journal}
  {\prd}\ }\textbf {\bibinfo {volume} {78}},\ \bibinfo {eid} {064036} (\bibinfo
  {year} {2008})},\ \Eprint {http://arxiv.org/abs/0806.4434} {arXiv:0806.4434
  [gr-qc]} \BibitemShut {NoStop}%
\bibitem [{\citenamefont {{Faraoni}}(2009)}]{Faraoni:2009rk}%
  \BibitemOpen
  \bibfield  {author} {\bibinfo {author} {\bibfnamefont {V.}~\bibnamefont
  {{Faraoni}}},\ }\bibfield  {title} {\enquote {\bibinfo {title} {{Lagrangian
  description of perfect fluids and modified gravity with an extra force}},}\
  }\href {\doibase 10.1103/PhysRevD.80.124040} {\bibfield  {journal} {\bibinfo
  {journal} {\prd}\ }\textbf {\bibinfo {volume} {80}},\ \bibinfo {eid} {124040}
  (\bibinfo {year} {2009})},\ \Eprint {http://arxiv.org/abs/0912.1249}
  {arXiv:0912.1249 [astro-ph.GA]} \BibitemShut {NoStop}%
\bibitem [{\citenamefont {Zeldovich}(1961)}]{zel61}%
  \BibitemOpen
  \bibfield  {author} {\bibinfo {author} {\bibfnamefont {Ya.B.}\ \bibnamefont
  {Zeldovich}},\ }\bibfield  {title} {\enquote {\bibinfo {title} {Equation of
  state at ultra-high densities and its relativistic limitations},}\
  }\href@noop {} {\bibfield  {journal} {\bibinfo  {journal} {Zhur. Eksptl'. i
  Teoret. Fiz}\ }\textbf {\bibinfo {volume} {41}} (\bibinfo {year}
  {1961})}\BibitemShut {NoStop}%
\bibitem [{\citenamefont {{Planck Collaboration}}\ \emph
  {et~al.}(2016)\citenamefont {{Planck Collaboration}}, \citenamefont {{Ade}},
  \citenamefont {{Aghanim}}, \citenamefont {{Arnaud}}, \citenamefont
  {{Ashdown}}, \citenamefont {{Aumont}}, \citenamefont {{Baccigalupi}},
  \citenamefont {{Banday}}, \citenamefont {{Barreiro}}, \citenamefont
  {{Bartlett}}, \citenamefont {{Bartolo}},\ and\ \citenamefont
  {et~al.}}]{Planck15Cosmo}%
  \BibitemOpen
  \bibfield  {author} {\bibinfo {author} {\bibnamefont {{Planck
  Collaboration}}}, \bibinfo {author} {\bibfnamefont {P.~A.~R.}\ \bibnamefont
  {{Ade}}}, \bibinfo {author} {\bibfnamefont {N.}~\bibnamefont {{Aghanim}}},
  \bibinfo {author} {\bibfnamefont {M.}~\bibnamefont {{Arnaud}}}, \bibinfo
  {author} {\bibfnamefont {M.}~\bibnamefont {{Ashdown}}}, \bibinfo {author}
  {\bibfnamefont {J.}~\bibnamefont {{Aumont}}}, \bibinfo {author}
  {\bibfnamefont {C.}~\bibnamefont {{Baccigalupi}}}, \bibinfo {author}
  {\bibfnamefont {A.~J.}\ \bibnamefont {{Banday}}}, \bibinfo {author}
  {\bibfnamefont {R.~B.}\ \bibnamefont {{Barreiro}}}, \bibinfo {author}
  {\bibfnamefont {J.~G.}\ \bibnamefont {{Bartlett}}}, \bibinfo {author}
  {\bibfnamefont {N.}~\bibnamefont {{Bartolo}}}, \ and\ \bibinfo {author}
  {\bibnamefont {et~al.}},\ }\bibfield  {title} {\enquote {\bibinfo {title}
  {{Planck 2015 results. XIII. Cosmological parameters}},}\ }\href {\doibase
  10.1051/0004-6361/201525830} {\bibfield  {journal} {\bibinfo  {journal}
  {\aap}\ }\textbf {\bibinfo {volume} {594}},\ \bibinfo {eid} {A13} (\bibinfo
  {year} {2016})},\ \Eprint {http://arxiv.org/abs/1502.01589}
  {arXiv:1502.01589} \BibitemShut {NoStop}%
\bibitem [{\citenamefont {Dodelson}(2003)}]{Dodelson:1282338}%
  \BibitemOpen
  \bibfield  {author} {\bibinfo {author} {\bibfnamefont {S.}~\bibnamefont
  {Dodelson}},\ }\href {https://cds.cern.ch/record/1282338} {\emph {\bibinfo
  {title} {{Modern cosmology}}}}\ (\bibinfo  {publisher} {Academic Press},\
  \bibinfo {address} {San Diego, CA},\ \bibinfo {year} {2003})\BibitemShut
  {NoStop}%
\bibitem [{\citenamefont {{Shapiro}}\ and\ \citenamefont
  {{Teukolsky}}(1983)}]{Shapiro83}%
  \BibitemOpen
  \bibfield  {author} {\bibinfo {author} {\bibfnamefont {S.~L.}\ \bibnamefont
  {{Shapiro}}}\ and\ \bibinfo {author} {\bibfnamefont {S.~A.}\ \bibnamefont
  {{Teukolsky}}},\ }\href@noop {} {\emph {\bibinfo {title} {Research supported
  by the National Science Foundation.~New York, Wiley-Interscience, 1983, 663
  p.}}}\ (\bibinfo {year} {1983})\BibitemShut {NoStop}%
\bibitem [{\citenamefont {{Press}}\ \emph {et~al.}(2002)\citenamefont
  {{Press}}, \citenamefont {{Teukolsky}}, \citenamefont {{Vetterling}},\ and\
  \citenamefont {{Flannery}}}]{numrec}%
  \BibitemOpen
  \bibfield  {author} {\bibinfo {author} {\bibfnamefont {W.~H.}\ \bibnamefont
  {{Press}}}, \bibinfo {author} {\bibfnamefont {S.~A.}\ \bibnamefont
  {{Teukolsky}}}, \bibinfo {author} {\bibfnamefont {W.~T.}\ \bibnamefont
  {{Vetterling}}}, \ and\ \bibinfo {author} {\bibfnamefont {B.~P.}\
  \bibnamefont {{Flannery}}},\ }\href@noop {} {\emph {\bibinfo {title}
  {Numerical recipes in C++ : the art of scientific computing by William
  H.~Press, Cambridge Univ. Press, Cambridge}}}\ (\bibinfo {year}
  {2002})\BibitemShut {NoStop}%
\bibitem [{\citenamefont {{Baym}}\ \emph {et~al.}(1971)\citenamefont {{Baym}},
  \citenamefont {{Pethick}},\ and\ \citenamefont {{Sutherland}}}]{bps}%
  \BibitemOpen
  \bibfield  {author} {\bibinfo {author} {\bibfnamefont {G.}~\bibnamefont
  {{Baym}}}, \bibinfo {author} {\bibfnamefont {C.}~\bibnamefont {{Pethick}}}, \
  and\ \bibinfo {author} {\bibfnamefont {P.}~\bibnamefont {{Sutherland}}},\
  }\bibfield  {title} {\enquote {\bibinfo {title} {{The Ground State of Matter
  at High Densities: Equation of State and Stellar Models}},}\ }\href {\doibase
  10.1086/151216} {\bibfield  {journal} {\bibinfo  {journal} {\apj}\ }\textbf
  {\bibinfo {volume} {170}},\ \bibinfo {pages} {299} (\bibinfo {year}
  {1971})}\BibitemShut {NoStop}%
\bibitem [{\citenamefont {Reinhard}\ \emph {et~al.}(1999)\citenamefont
  {Reinhard}, \citenamefont {Dean}, \citenamefont {Nazarewicz}, \citenamefont
  {Dobaczewski}, \citenamefont {Maruhn},\ and\ \citenamefont
  {Strayer}}]{Rei99}%
  \BibitemOpen
  \bibfield  {author} {\bibinfo {author} {\bibfnamefont {P.-G.}\ \bibnamefont
  {Reinhard}}, \bibinfo {author} {\bibfnamefont {D.~J.}\ \bibnamefont {Dean}},
  \bibinfo {author} {\bibfnamefont {W.}~\bibnamefont {Nazarewicz}}, \bibinfo
  {author} {\bibfnamefont {J.}~\bibnamefont {Dobaczewski}}, \bibinfo {author}
  {\bibfnamefont {J.~A.}\ \bibnamefont {Maruhn}}, \ and\ \bibinfo {author}
  {\bibfnamefont {M.~R.}\ \bibnamefont {Strayer}},\ }\bibfield  {title}
  {\enquote {\bibinfo {title} {Shape coexistence and the effective
  nucleon-nucleon interaction},}\ }\href {\doibase 10.1103/PhysRevC.60.014316}
  {\bibfield  {journal} {\bibinfo  {journal} {Phys. Rev. C}\ }\textbf {\bibinfo
  {volume} {60}},\ \bibinfo {pages} {014316} (\bibinfo {year}
  {1999})}\BibitemShut {NoStop}%
\bibitem [{\citenamefont {Danielewicz}\ and\ \citenamefont
  {Lee}(2009)}]{Dan09}%
  \BibitemOpen
  \bibfield  {author} {\bibinfo {author} {\bibfnamefont {P.}~\bibnamefont
  {Danielewicz}}\ and\ \bibinfo {author} {\bibfnamefont {J.}~\bibnamefont
  {Lee}},\ }\bibfield  {title} {\enquote {\bibinfo {title} {Symmetry energy $i$
  : Semi-infinite matter},}\ }\href {\doibase
  https://doi.org/10.1016/j.nuclphysa.2008.11.007} {\bibfield  {journal}
  {\bibinfo  {journal} {Nuclear Physics A}\ }\textbf {\bibinfo {volume}
  {818}},\ \bibinfo {pages} {36 -- 96} (\bibinfo {year} {2009})}\BibitemShut
  {NoStop}%
\bibitem [{\citenamefont {Gulminelli}\ and\ \citenamefont
  {Raduta}(2015)}]{GR15}%
  \BibitemOpen
  \bibfield  {author} {\bibinfo {author} {\bibfnamefont {F.}~\bibnamefont
  {Gulminelli}}\ and\ \bibinfo {author} {\bibfnamefont {Ad.~R.}\ \bibnamefont
  {Raduta}},\ }\bibfield  {title} {\enquote {\bibinfo {title} {Unified
  treatment of subsaturation stellar matter at zero and finite temperature},}\
  }\href {\doibase 10.1103/PhysRevC.92.055803} {\bibfield  {journal} {\bibinfo
  {journal} {Phys. Rev. C}\ }\textbf {\bibinfo {volume} {92}},\ \bibinfo
  {pages} {055803} (\bibinfo {year} {2015})}\BibitemShut {NoStop}%
\bibitem [{\citenamefont {{M{\"u}ller}}\ and\ \citenamefont
  {{Serot}}(1996)}]{ref_ms2}%
  \BibitemOpen
  \bibfield  {author} {\bibinfo {author} {\bibfnamefont {H.}~\bibnamefont
  {{M{\"u}ller}}}\ and\ \bibinfo {author} {\bibfnamefont {B.~D.}\ \bibnamefont
  {{Serot}}},\ }\bibfield  {title} {\enquote {\bibinfo {title} {{Relativistic
  mean-field theory and the high-density nuclear equation of state}},}\ }\href
  {\doibase 10.1016/0375-9474(96)00187-X} {\bibfield  {journal} {\bibinfo
  {journal} {Nuclear Physics A}\ }\textbf {\bibinfo {volume} {606}},\ \bibinfo
  {pages} {508--537} (\bibinfo {year} {1996})},\ \Eprint
  {http://arxiv.org/abs/arXiv:nucl-th/9603037} {arXiv:nucl-th/9603037}
  \BibitemShut {NoStop}%
\bibitem [{\citenamefont {Akmal}\ \emph {et~al.}(1998)\citenamefont {Akmal},
  \citenamefont {Pandharipande},\ and\ \citenamefont {Ravenhall}}]{ref_APR}%
  \BibitemOpen
  \bibfield  {author} {\bibinfo {author} {\bibfnamefont {A.}~\bibnamefont
  {Akmal}}, \bibinfo {author} {\bibfnamefont {V.~R.}\ \bibnamefont
  {Pandharipande}}, \ and\ \bibinfo {author} {\bibfnamefont {D.~G.}\
  \bibnamefont {Ravenhall}},\ }\bibfield  {title} {\enquote {\bibinfo {title}
  {Equation of state of nucleon matter and neutron star structure},}\ }\href
  {\doibase 10.1103/PhysRevC.58.1804} {\bibfield  {journal} {\bibinfo
  {journal} {Phys. Rev. C}\ }\textbf {\bibinfo {volume} {58}},\ \bibinfo
  {pages} {1804--1828} (\bibinfo {year} {1998})}\BibitemShut {NoStop}%
\bibitem [{\citenamefont {{Douchin}}\ and\ \citenamefont
  {{Haensel}}(2001)}]{ref_SLY}%
  \BibitemOpen
  \bibfield  {author} {\bibinfo {author} {\bibfnamefont {F.}~\bibnamefont
  {{Douchin}}}\ and\ \bibinfo {author} {\bibfnamefont {P.}~\bibnamefont
  {{Haensel}}},\ }\bibfield  {title} {\enquote {\bibinfo {title} {{A unified
  equation of state of dense matter and neutron star structure}},}\ }\href
  {\doibase 10.1051/0004-6361:20011402} {\bibfield  {journal} {\bibinfo
  {journal} {\aap}\ }\textbf {\bibinfo {volume} {380}},\ \bibinfo {pages}
  {151--167} (\bibinfo {year} {2001})},\ \Eprint
  {http://arxiv.org/abs/arXiv:astro-ph/0111092} {arXiv:astro-ph/0111092}
  \BibitemShut {NoStop}%
\bibitem [{\citenamefont {Glendenning}\ and\ \citenamefont
  {Moszkowski}(1991)}]{ref_GM1}%
  \BibitemOpen
  \bibfield  {author} {\bibinfo {author} {\bibfnamefont {N.~K.}\ \bibnamefont
  {Glendenning}}\ and\ \bibinfo {author} {\bibfnamefont {S.~A.}\ \bibnamefont
  {Moszkowski}},\ }\bibfield  {title} {\enquote {\bibinfo {title}
  {Reconciliation of neutron-star masses and binding of the
  \ensuremath{\Lambda} in hypernuclei},}\ }\href {\doibase
  10.1103/PhysRevLett.67.2414} {\bibfield  {journal} {\bibinfo  {journal}
  {Phys. Rev. Lett.}\ }\textbf {\bibinfo {volume} {67}},\ \bibinfo {pages}
  {2414--2417} (\bibinfo {year} {1991})}\BibitemShut {NoStop}%
\bibitem [{\citenamefont {Oertel}\ \emph {et~al.}(2015)\citenamefont {Oertel},
  \citenamefont {Providência}, \citenamefont {Gulminelli},\ and\ \citenamefont
  {Raduta}}]{Oer+15}%
  \BibitemOpen
  \bibfield  {author} {\bibinfo {author} {\bibfnamefont {M}~\bibnamefont
  {Oertel}}, \bibinfo {author} {\bibfnamefont {C}~\bibnamefont {Providência}},
  \bibinfo {author} {\bibfnamefont {F}~\bibnamefont {Gulminelli}}, \ and\
  \bibinfo {author} {\bibfnamefont {Ad~R}\ \bibnamefont {Raduta}},\ }\bibfield
  {title} {\enquote {\bibinfo {title} {Hyperons in neutron star matter within
  relativistic mean-field models},}\ }\href
  {http://stacks.iop.org/0954-3899/42/i=7/a=075202} {\bibfield  {journal}
  {\bibinfo  {journal} {Journal of Physics G: Nuclear and Particle Physics}\
  }\textbf {\bibinfo {volume} {42}},\ \bibinfo {pages} {075202} (\bibinfo
  {year} {2015})}\BibitemShut {NoStop}%
\bibitem [{\citenamefont {{Lattimer}}\ and\ \citenamefont
  {{Prakash}}(2001)}]{lat01}%
  \BibitemOpen
  \bibfield  {author} {\bibinfo {author} {\bibfnamefont {J.~M.}\ \bibnamefont
  {{Lattimer}}}\ and\ \bibinfo {author} {\bibfnamefont {M.}~\bibnamefont
  {{Prakash}}},\ }\bibfield  {title} {\enquote {\bibinfo {title} {{Neutron Star
  Structure and the Equation of State}},}\ }\href {\doibase 10.1086/319702}
  {\bibfield  {journal} {\bibinfo  {journal} {\apj}\ }\textbf {\bibinfo
  {volume} {550}},\ \bibinfo {pages} {426--442} (\bibinfo {year} {2001})},\
  \Eprint {http://arxiv.org/abs/arXiv:astro-ph/0002232}
  {arXiv:astro-ph/0002232} \BibitemShut {NoStop}%
\bibitem [{\citenamefont {{Ek{\c s}i}}(2016)}]{eks16}%
  \BibitemOpen
  \bibfield  {author} {\bibinfo {author} {\bibfnamefont {K.~Y.}\ \bibnamefont
  {{Ek{\c s}i}}},\ }\bibfield  {title} {\enquote {\bibinfo {title} {{Neutron
  stars: compact objects with relativistic gravity}},}\ }\href {\doibase
  10.3906/fiz-1510-11} {\bibfield  {journal} {\bibinfo  {journal} {Turkish J.\
  of Phys.}\ }\textbf {\bibinfo {volume} {40}},\ \bibinfo {pages} {127--138}
  (\bibinfo {year} {2016})},\ \Eprint {http://arxiv.org/abs/1511.04305}
  {arXiv:1511.04305 [astro-ph.HE]} \BibitemShut {NoStop}%
\bibitem [{\citenamefont {{He}}\ \emph {et~al.}(2015)\citenamefont {{He}},
  \citenamefont {{Fattoyev}}, \citenamefont {{Li}},\ and\ \citenamefont
  {{Newton}}}]{he+15}%
  \BibitemOpen
  \bibfield  {author} {\bibinfo {author} {\bibfnamefont {X.-T.}\ \bibnamefont
  {{He}}}, \bibinfo {author} {\bibfnamefont {F.~J.}\ \bibnamefont
  {{Fattoyev}}}, \bibinfo {author} {\bibfnamefont {B.-A.}\ \bibnamefont
  {{Li}}}, \ and\ \bibinfo {author} {\bibfnamefont {W.~G.}\ \bibnamefont
  {{Newton}}},\ }\bibfield  {title} {\enquote {\bibinfo {title} {{Impact of the
  equation-of-state-gravity degeneracy on constraining the nuclear symmetry
  energy from astrophysical observables}},}\ }\href {\doibase
  10.1103/PhysRevC.91.015810} {\bibfield  {journal} {\bibinfo  {journal}
  {\prc}\ }\textbf {\bibinfo {volume} {91}},\ \bibinfo {eid} {015810} (\bibinfo
  {year} {2015})},\ \Eprint {http://arxiv.org/abs/1408.0857} {arXiv:1408.0857
  [nucl-th]} \BibitemShut {NoStop}%
\bibitem [{\citenamefont {{{\"O}zel}}\ and\ \citenamefont
  {{Freire}}(2016)}]{oze16}%
  \BibitemOpen
  \bibfield  {author} {\bibinfo {author} {\bibfnamefont {F.}~\bibnamefont
  {{{\"O}zel}}}\ and\ \bibinfo {author} {\bibfnamefont {P.}~\bibnamefont
  {{Freire}}},\ }\bibfield  {title} {\enquote {\bibinfo {title} {{Masses,
  Radii, and the Equation of State of Neutron Stars}},}\ }\href {\doibase
  10.1146/annurev-astro-081915-023322} {\bibfield  {journal} {\bibinfo
  {journal} {\araa}\ }\textbf {\bibinfo {volume} {54}},\ \bibinfo {pages}
  {401--440} (\bibinfo {year} {2016})},\ \Eprint
  {http://arxiv.org/abs/1603.02698} {arXiv:1603.02698 [astro-ph.HE]}
  \BibitemShut {NoStop}%
\bibitem [{\citenamefont {{Harrison}}\ \emph {et~al.}(1965)\citenamefont
  {{Harrison}}, \citenamefont {{Thorne}}, \citenamefont {{Wakano}},\ and\
  \citenamefont {{Wheeler}}}]{har+65}%
  \BibitemOpen
  \bibfield  {author} {\bibinfo {author} {\bibfnamefont {B.~K.}\ \bibnamefont
  {{Harrison}}}, \bibinfo {author} {\bibfnamefont {K.~S.}\ \bibnamefont
  {{Thorne}}}, \bibinfo {author} {\bibfnamefont {M.}~\bibnamefont {{Wakano}}},
  \ and\ \bibinfo {author} {\bibfnamefont {J.~A.}\ \bibnamefont {{Wheeler}}},\
  }\href@noop {} {\emph {\bibinfo {title} {Gravitation Theory and Gravitational
  Collapse, Chicago: University of Chicago Press, 1965}}}\ (\bibinfo {year}
  {1965})\BibitemShut {NoStop}%
\bibitem [{\citenamefont {{Chavanis}}(2015)}]{Chavanis15}%
  \BibitemOpen
  \bibfield  {author} {\bibinfo {author} {\bibfnamefont {P.-H.}\ \bibnamefont
  {{Chavanis}}},\ }\bibfield  {title} {\enquote {\bibinfo {title} {{Cosmology
  with a stiff matter era}},}\ }\href {\doibase 10.1103/PhysRevD.92.103004}
  {\bibfield  {journal} {\bibinfo  {journal} {\prd}\ }\textbf {\bibinfo
  {volume} {92}},\ \bibinfo {eid} {103004} (\bibinfo {year} {2015})},\ \Eprint
  {http://arxiv.org/abs/1412.0743} {arXiv:1412.0743 [gr-qc]} \BibitemShut
  {NoStop}%
\bibitem [{\citenamefont {{Chandrasekhar}}(1964a)}]{cha64a}%
  \BibitemOpen
  \bibfield  {author} {\bibinfo {author} {\bibfnamefont {S.}~\bibnamefont
  {{Chandrasekhar}}},\ }\bibfield  {title} {\enquote {\bibinfo {title} {{The
  Dynamical Instability of Gaseous Masses Approaching the Schwarzschild Limit
  in General Relativity.}}}\ }\href {\doibase 10.1086/147938} {\bibfield
  {journal} {\bibinfo  {journal} {\apj}\ }\textbf {\bibinfo {volume} {140}},\
  \bibinfo {pages} {417} (\bibinfo {year} {1964a})}\BibitemShut {NoStop}%
\bibitem [{\citenamefont {{Chandrasekhar}}(1964b)}]{cha64b}%
  \BibitemOpen
  \bibfield  {author} {\bibinfo {author} {\bibfnamefont {S.}~\bibnamefont
  {{Chandrasekhar}}},\ }\bibfield  {title} {\enquote {\bibinfo {title}
  {{Erratum: the Dynamical Instability of Gaseous Masses Approaching the
  Schwarzschild Limit in General Relativity.}}}\ }\href {\doibase
  10.1086/148040} {\bibfield  {journal} {\bibinfo  {journal} {\apj}\ }\textbf
  {\bibinfo {volume} {140}},\ \bibinfo {pages} {1342} (\bibinfo {year}
  {1964b})}\BibitemShut {NoStop}%
\bibitem [{\citenamefont {{Datta}}\ \emph {et~al.}(1992)\citenamefont
  {{Datta}}, \citenamefont {{Sahu}}, \citenamefont {{Anand}},\ and\
  \citenamefont {{Goyal}}}]{dat+92}%
  \BibitemOpen
  \bibfield  {author} {\bibinfo {author} {\bibfnamefont {B.}~\bibnamefont
  {{Datta}}}, \bibinfo {author} {\bibfnamefont {P.~K.}\ \bibnamefont {{Sahu}}},
  \bibinfo {author} {\bibfnamefont {J.~D.}\ \bibnamefont {{Anand}}}, \ and\
  \bibinfo {author} {\bibfnamefont {A.}~\bibnamefont {{Goyal}}},\ }\bibfield
  {title} {\enquote {\bibinfo {title} {{Eigenfrequencies of radial pulsations
  of strange quark stars}},}\ }\href {\doibase 10.1016/0370-2693(92)90025-Y}
  {\bibfield  {journal} {\bibinfo  {journal} {Physics Letters B}\ }\textbf
  {\bibinfo {volume} {283}},\ \bibinfo {pages} {313--318} (\bibinfo {year}
  {1992})},\ \Eprint {http://arxiv.org/abs/astro-ph/9304024} {astro-ph/9304024}
  \BibitemShut {NoStop}%
\bibitem [{\citenamefont {{Haensel}}\ \emph {et~al.}(2007)\citenamefont
  {{Haensel}}, \citenamefont {{Potekhin}},\ and\ \citenamefont
  {{Yakovlev}}}]{han+07}%
  \BibitemOpen
  \bibinfo {editor} {\bibfnamefont {P.}~\bibnamefont {{Haensel}}}, \bibinfo
  {editor} {\bibfnamefont {A.~Y.}\ \bibnamefont {{Potekhin}}}, \ and\ \bibinfo
  {editor} {\bibfnamefont {D.~G.}\ \bibnamefont {{Yakovlev}}},\ eds.,\
  \href@noop {} {\emph {\bibinfo {title} {Astrophysics and Space Science
  Library}}},\ \bibinfo {series} {Astrophysics and Space Science Library, New
  York Springer}, Vol.\ \bibinfo {volume} {326}\ (\bibinfo {year}
  {2007})\BibitemShut {NoStop}%
\end{thebibliography}%

\end{document}